\documentclass[useAMS,usenatbib]{mn2e}
\usepackage{graphicx}

\title[A new benchmark T8-9 brown dwarf]{A new benchmark T8-9 brown dwarf and a couple of new mid-T dwarfs from the UKIDSS DR5+ LAS\thanks{Based on observations collected at the German-Spanish Astronomical Center, Calar Alto, jointly operated by the Max-Planck-Institut f{\"u}r Astronomie Heidelberg and the  Instituto de Astrof{\'i}sica de Andaluc'a (CSIC); and on observations made with ESO/MPG Telescope at the La Silla  Observatory under programme ID 081.A-9012 and 081.A-9014.}}
\author[B. Goldman et al.]{
B. Goldman$^1$\thanks{E-mail: goldman@mpia.de},
S. Marsat$^{1}$,
T. Henning$^{1}$,
C. Clemens$^2$,
and J. Greiner$^2$
\\
$^{1}$ Max Planck Institute for Astronomy, Koenigstuhl 17, D--69117 Heidelberg, Germany \\
$^2$ Max Planck Institute for extraterrestrial Physics,
Giessenbachstra{\ss}e, D--85748 Garching, Germany
}
\begin{document}
%
%
%
%


\def\aj{\rm{AJ}}                   
\def\araa{\rm{ARA\&A}}             
\def\apj{\rm{ApJ}}                 
\def\apjl{\rm{ApJ}}                
\def\apjs{\rm{ApJS}}               
\def\ao{\rm{Appl.~Opt.}}           
\def\apss{\rm{Ap\&SS}}             
\def\aap{\rm{A\&A}}                
\def\aapr{\rm{A\&A~Rev.}}          
\def\aaps{\rm{A\&AS}}              
\def\azh{\rm{AZh}}                 
\def\baas{\rm{BAAS}}               
\def\jrasc{\rm{JRASC}}             
\def\memras{\rm{MmRAS}}            
\def\mnras{\rm{MNRAS}}             
\def\pra{\rm{Phys.~Rev.~A}}        
\def\prb{\rm{Phys.~Rev.~B}}        
\def\prc{\rm{Phys.~Rev.~C}}        
\def\prd{\rm{Phys.~Rev.~D}}        
\def\pre{\rm{Phys.~Rev.~E}}        
\def\prl{\rm{Phys.~Rev.~Lett.}}    
\def\pasp{\rm{PASP}}               
\def\pasj{\rm{PASJ}}               
\def\qjras{\rm{QJRAS}}             
\def\skytel{\rm{S\&T}}             
\def\solphys{\rm{Sol.~Phys.}}      
\def\sovast{\rm{Soviet~Ast.}}      
\def\ssr{\rm{Space~Sci.~Rev.}}     
\def\zap{\rm{ZAp}}                 
\def\nat{\rm{Nature}}              
\def\iaucirc{\rm{IAU~Circ.}}       
\def\aplett{\rm{Astrophys.~Lett.}} 
\def\apspr{\rm{Astrophys.~Space~Phys.~Res.}}
\def\bain{\rm{Bull.~Astron.~Inst.~Netherlands}} 
\def\fcp{\rm{Fund.~Cosmic~Phys.}}  
\def\gca{\rm{Geochim.~Cosmochim.~Acta}}   
\def\grl{\rm{Geophys.~Res.~Lett.}} 
\def\jcp{\rm{J.~Chem.~Phys.}}      
\def\jgr{\rm{J.~Geophys.~Res.}}    
\def\jqsrt{\rm{J.~Quant.~Spec.~Radiat.~Transf.}}
\def\memsai{\rm{Mem.~Soc.~Astron.~Italiana}}
\def\nphysa{\rm{Nucl.~Phys.~A}}   
\def\physrep{\rm{Phys.~Rep.}}   
\def\physscr{\rm{Phys.~Scr}}   
\def\planss{\rm{Planet.~Space~Sci.}}   
\def\procspie{\rm{Proc.~SPIE}}   

\let\astap=\aap
\let\apjlett=\apjl
\let\apjsupp=\apjs
\let\applopt=\ao

\date{ }

\pagerange{\pageref{firstpage}--\pageref{lastpage}} \pubyear{2009}

\maketitle

\label{firstpage}

\begin{abstract}
Benchmark brown dwarfs are those {objects for which fiducial constraints are available}, including effective temperature, parallax, age, metallicity. 
We searched for new cool brown dwarfs in 186\,sq.deg of the {new} area covered by the data release DR5+ of the UKIDSS Large Area Survey. 
Follow-up optical and near-infrared broad-band photometry, and methane imaging of four promising candidates, revealed three objects with distinct methane absorption, typical of mid- to late-T dwarfs, and one possibly T4 dwarf.
The latest-type object, classified as T8--9, shares its large proper motion with Ross~458 (BD+13$^{\rm o}$2618), an active M0.5 binary which is 102\arcsec\ away, forming a hierarchical low-mass star+brown dwarf system.
Ross~458C has an absolute $J$-band magnitude of 16.4, and seems overluminous, particularly in the $K$ band, compared to similar field brown dwarfs.
We estimate the age of the system to be less than 1\,Gyr, and its mass to be as low as 14~Jupiter masses for the age of 1\,Gyr.
At 11.4\,pc, this new late~T~benchmark dwarf is a promising target to constrain the evolutionary and atmospheric models of very low-mass brown dwarfs.
{We present proper motion measurements for our targets and for 13 known brown dwarfs. 
Two brown dwarfs have velocities typical of the thick disk and may be old brown dwarfs.
} 
\end{abstract}

\begin{keywords}
stars: low-mass, brown dwarfs -- binaries: general -- stars: individual: Ross~458C.
\end{keywords}

\section{Introduction}

Cool brown dwarfs with effective temperatures lower than 1800\,K are characterized by complicated and diverse atmospheric processes that profoundly affect their emerging spectral energy distribution \citep[e.g.][]{Saumo08}. 
Atmospheric models have been much improved over the last decade \citep{Allar01,Tsuji05,Burro06,Saumo08,Helli08cm}.
Dynamical processes are investigated through observations \citep{Legge09} and simulations \citep{Helli08fo}.
Increasing the searched volume may reveal peculiar objects in terms of colours, metallicity and age \citep[e.g.][]{Burga04sd,Kirkp08} and lower effective temperatures. 

A valuable sub-class of brown dwarfs are those for which we can obtain more information than is usually the case for isolated brown dwarfs. As brown dwarfs cool with age, it is difficult to disentangle the effects of an higher mass and an younger age; gravity measurements are difficult to obtain for those faint objects, and  high-spectral resolution modeling is complex. Objects associated with a cluster, or a brighter companion, especially when they are close enough to allow high signal-to-noise spectroscopy, may be used as {\it benchmark brown dwarfs} \citep{Pinfi06,Burni09}.

{There are nine such systems including a T-type companion.
\citet{Legge02}, for instance, have studied the first two field L- and T-type brown dwarfs companions to main sequence stars, the T dwarf Gl~229B \citep{Nakaj95} and the L dwarf LHS~102B \citep{Goldm99}, to derive the substellar parameters of those objects and confirm the evolutionary models. 
Very recently, \citet{Faher10} presented a M3+T6.5 binary, G~204-39 and SDSS J1758+4633, with an age of 0.5--1.5\,Gyr, as well as one of the weakliest bound system, a G5+L4 binary with a 25,000-AU projected separation, G~200-28 and SDSS J1416+5006.
Finally, \citet{Burni09} found a M4+T8.5 system in the UKIDSS LAS DR~4, Wolf~940~AB, whose primary's low activity level indicates an age of 3.5--6\,Gyr.
}

Recent surveys, such as 2MASS, DENIS or SDSS
have brought a wealth of new neighbours of always cooler temperatures \citep{Delfo99,Kirkp00,Legge00}. 
The UKIDSS set of surveys, with a gain in volume of two dozens,  
push this research forward
\citep{Lawre07}. In its high-galactic latitude, shallow component, the Large Area Survey (LAS), more than 30 new T~dwarfs have been discovered, including the coolest compact objects outside the Solar system \citep{Lodie07,Warre07,Chiu08, Pinfi08,Burni08}, along with the CFHT-LS survey \citep{Delor08}.

We have searched the area newly covered by the data release DR5+ of April~2009 of UKIDSS--LAS for new cool brown dwarfs of spectral type~T. Using a simple colour-cut selection, we selected targets for photometric follow-up. Here, we report on the discovery of a new benchmark T8--9 brown dwarf located at 11\,pc, and of 3--4 new mid-T brown dwarfs.

To obtain a preliminary spectral classification, we rely on methane imaging.
Spectra of mid-T dwarfs and later-type brown dwarfs are known to exhibit strong methane (and water) absorption in the red half of the $H$ band. This feature distinguishes them from all celestial bodies warmer than 1400\,K. Relatively short exposure times in the 0.1-$\umu$m-wide filters offer a robust confirmation and a preliminary spectral classification of brown dwarf candidates (see Appendix\,\ref{methane_cal}).
{This method has been successfully used by \citet{Tinne05} to confirm candidates found in 2MASS  and \citet{Lodie09}, in the UKIDSS  Deep Extragalactic Survey.}

We establish the companionship of ULAS~J130041+122114 with Ross~458 (BD+13$^{\rm o}$2618, GJ~494, HIP~63510) in Section\,\ref{companion}, and refer to it as Ross~458C for simplicity. \citet{Ross26} discovered the proper motion of Ross~458.
His work allows us to establish the companionship with our target, and we therefore choose this name among all the possible names of Ross~458.

We first describe our selection criteria and follow-up observations and data reduction. 
In Section\,\ref{companion} we discuss the companionship of Ross~458C with Ross~458AB. 
Finally, in Section\,\ref{prop} we study the properties of the new brown dwarfs and brown dwarf candidates.
The calibration of the methane filter set of Omega~2000 is detailed in the Appendix\,\ref{methane_cal}.

\section[]{Candidate selection and follow-up observations}

\subsection{Mining the UKIDSS DR5+ of the LAS}

In this subsection we discuss how we selected the four objects for follow-up observations. We do not intend in this article to search for cool brown dwarfs systematically and we refer to a later article (and a larger sample) for a proper detection selection and efficiency estimate.

\begin{figure}
\includegraphics[width=.48\textwidth]{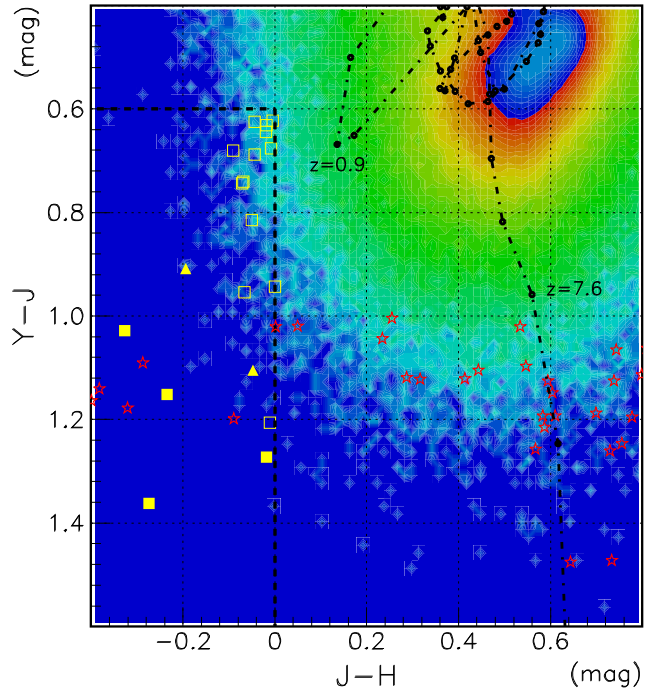}
\caption{$Y-J$ vs. $J-H$ colour-colour diagramme for UKIDSS DR5+. The selection cuts are indicated by the dashed lines. The targets described in this {paper} are shown as full squares, while the empty squares indicate objects with an SDSS cross-match. The two triangles indicate good targets for which we could not obtain methane imaging data. 
The grey shades scale as the logarithm of the stellar density, with black for the highest density of the stellar locus, selected {over the whole area of} DR5+ with the same quality cuts as our candidate sample. Empty stars show known brown dwarfs; the dash-dotted line shows the predicted quasar locus, with redshift steps of 0.1 \citep{Hewet06}.
}
\label{Fselec}
\end{figure}

Known cool brown dwarfs (mid-T type and later) are red in their optical and $Y-J$ colour and blue in their near-IR colours. The former is primarily due to the absorption by alkali elements such as Na\,I and K\,I, while the latter is due to the absorption by water and methane.
This behaviour is predicted by models \citep{Burro03}, and confirmed by observations of 2MASS and SDSS brown dwarfs \citep{Hewet06}, as well as recent UKIDSS discoveries \citep{Warre07,Lodie07,Pinfi08}. \citet{Burni08} searched for $Y-J>0.5$ and $J-H<0.1$ objects and the finding closest to these colour cuts (in a sample of three) has UKIDSS colours of $Y-J=0.60\pm0.07$ and \mbox{$J-H=-0.25\pm0.03$}.
Fig.\,\ref{Fselec} summarizes these results.

We search the UKIDSS DR5+ data release of the LAS  for objects with $YJH$ detections and photometric errors smaller than 30\,mmag; $YJ$ {\tt ppErrBits} quality flags smaller than 256; a $J$-band magnitude between 15.0 and 20.0\,mag; a probability to be stellar larger than 0.6 and a class star parameter {\tt mergedClass} of $-1$. The objects should fulfill the following colour constraints (using  2\arcsec-radius aperture photometry): $Y-J>0.6$\,mag, $J-H<0.0$\,mag and either $H-K<0.25$\,mag if the object is detected in the $K$ band, or $H-{\rm depth}_K<0.25$ where ${\rm depth}_K$ is the limiting magnitude (3$\sigma)$ in a 2\arcsec-radius aperture.
In order to avoid duplicating observations, we perform the same search on the DR4+ LAS data release and exclude from the DR5+ results the objects found identical in the DR4+ results.

\begin{figure}
\includegraphics[width=.235\textwidth]{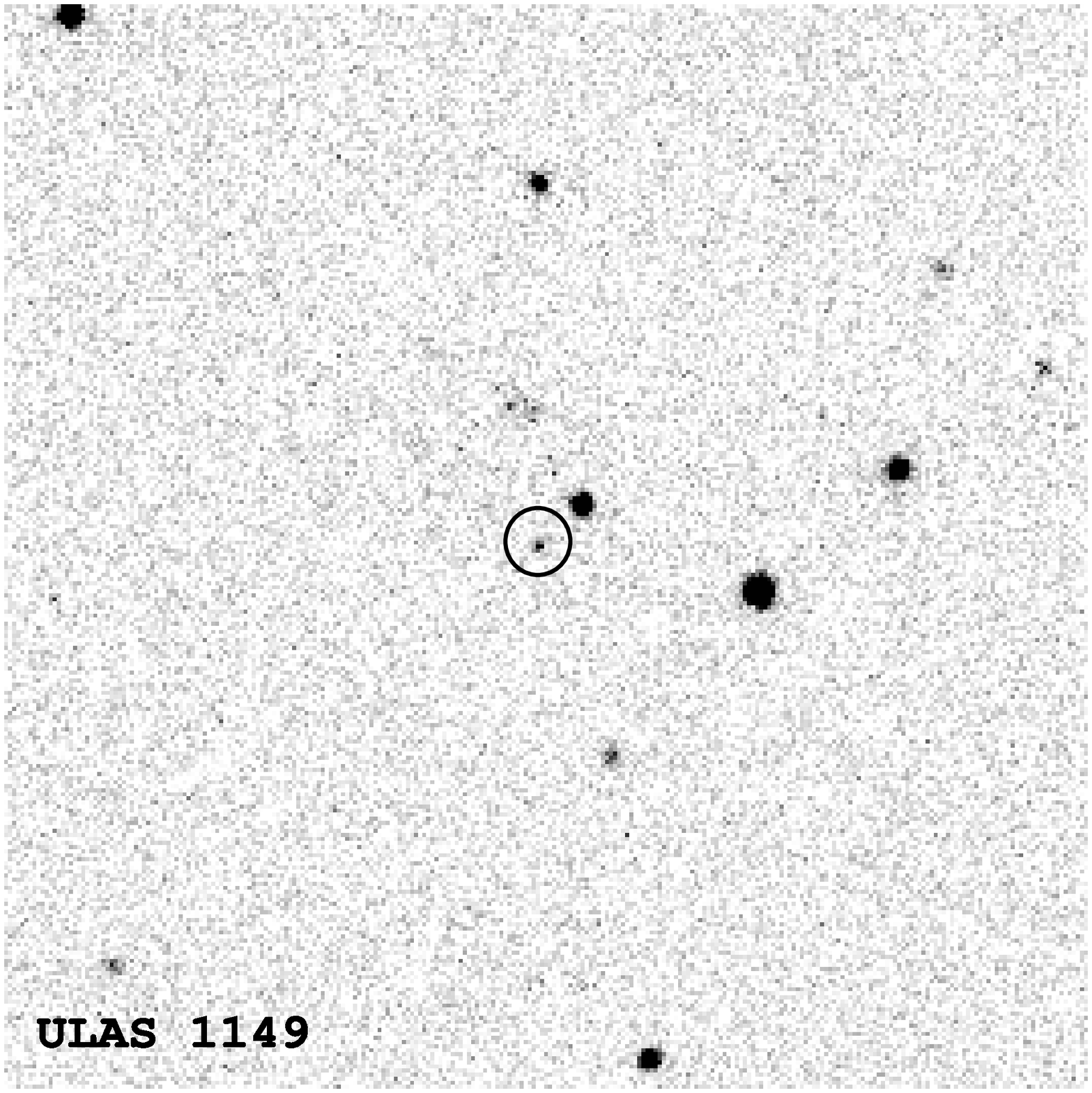}
\includegraphics[width=.235\textwidth]{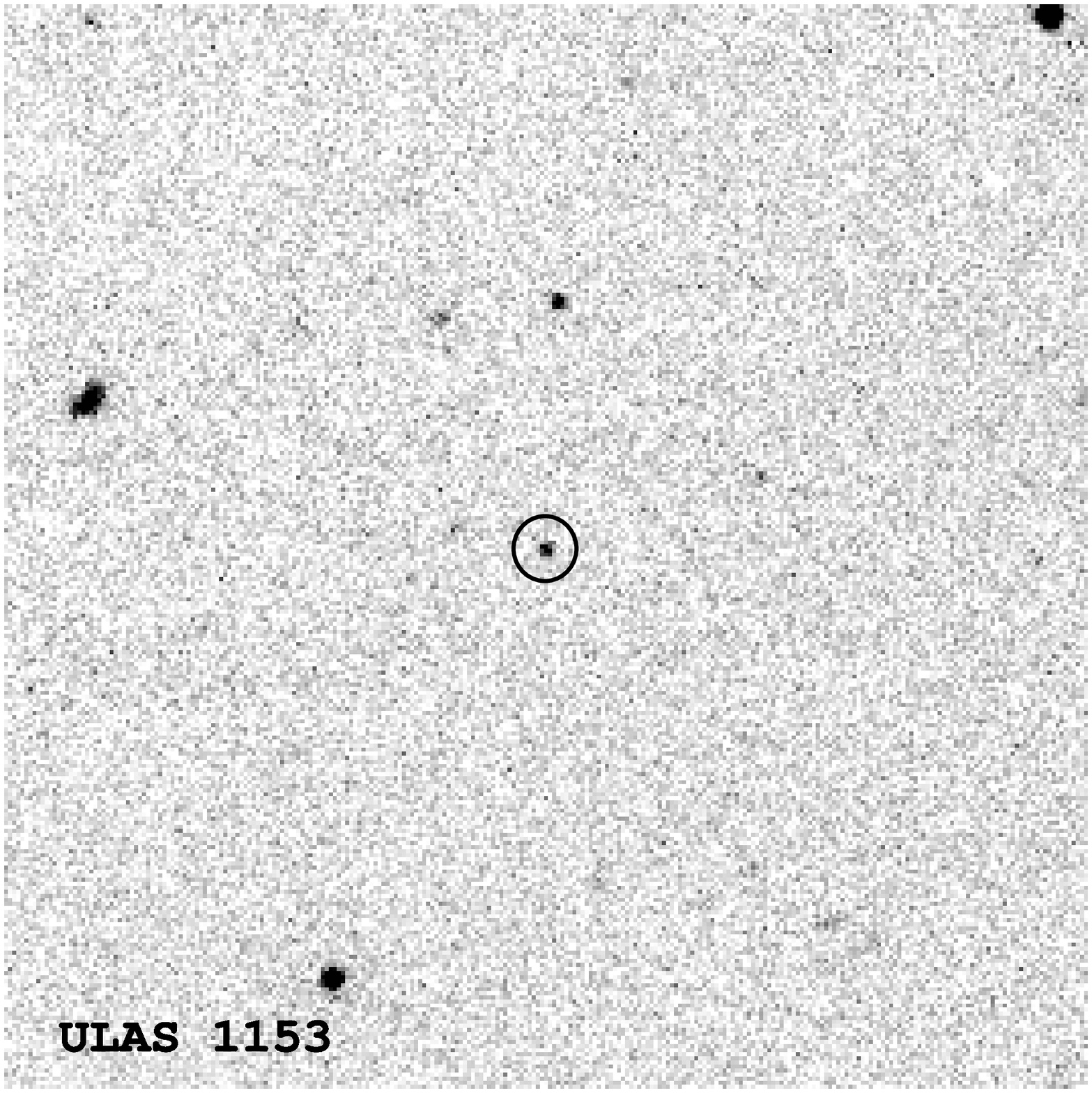}
\includegraphics[width=.235\textwidth]{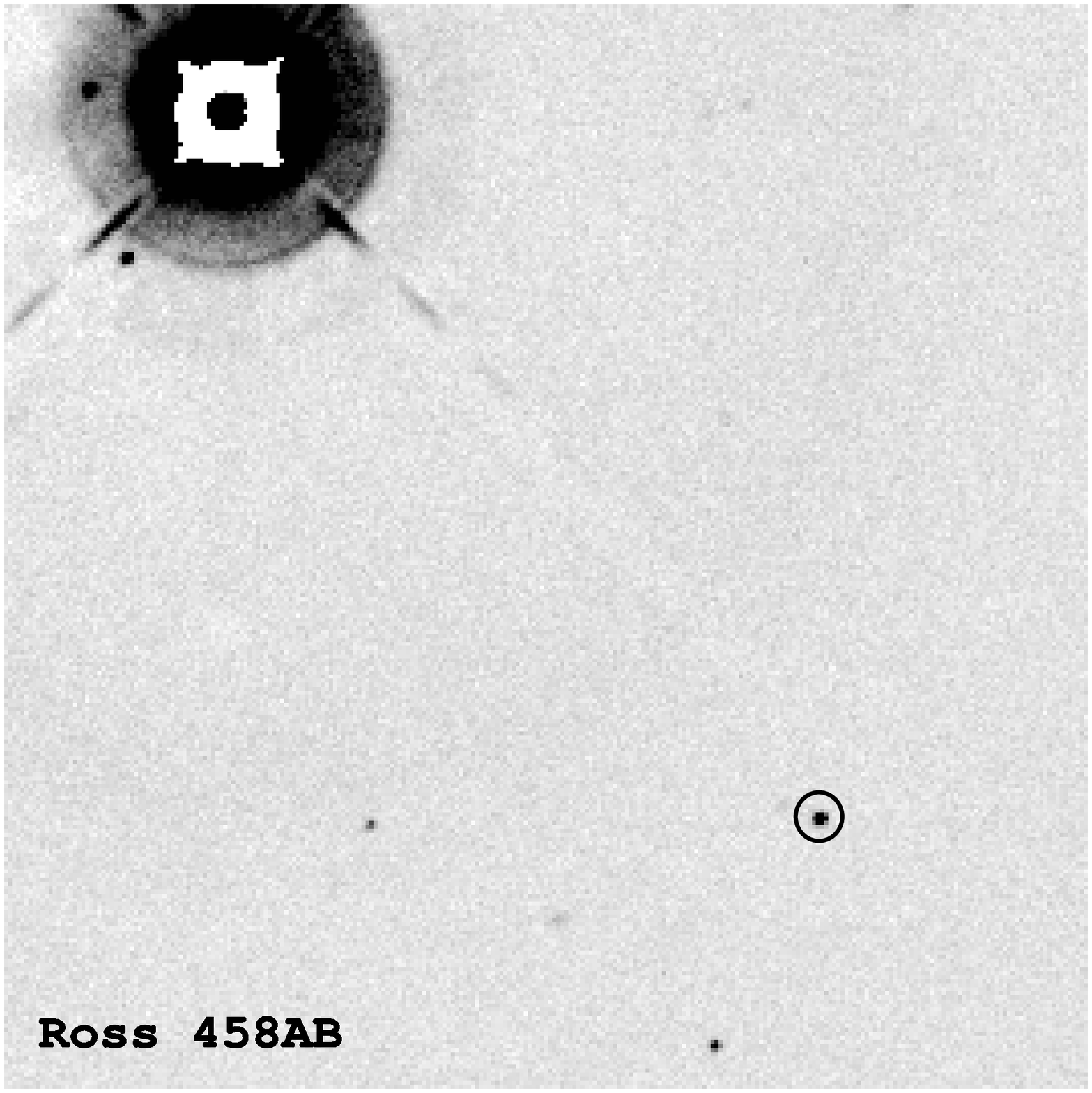} 
\includegraphics[width=.235\textwidth]{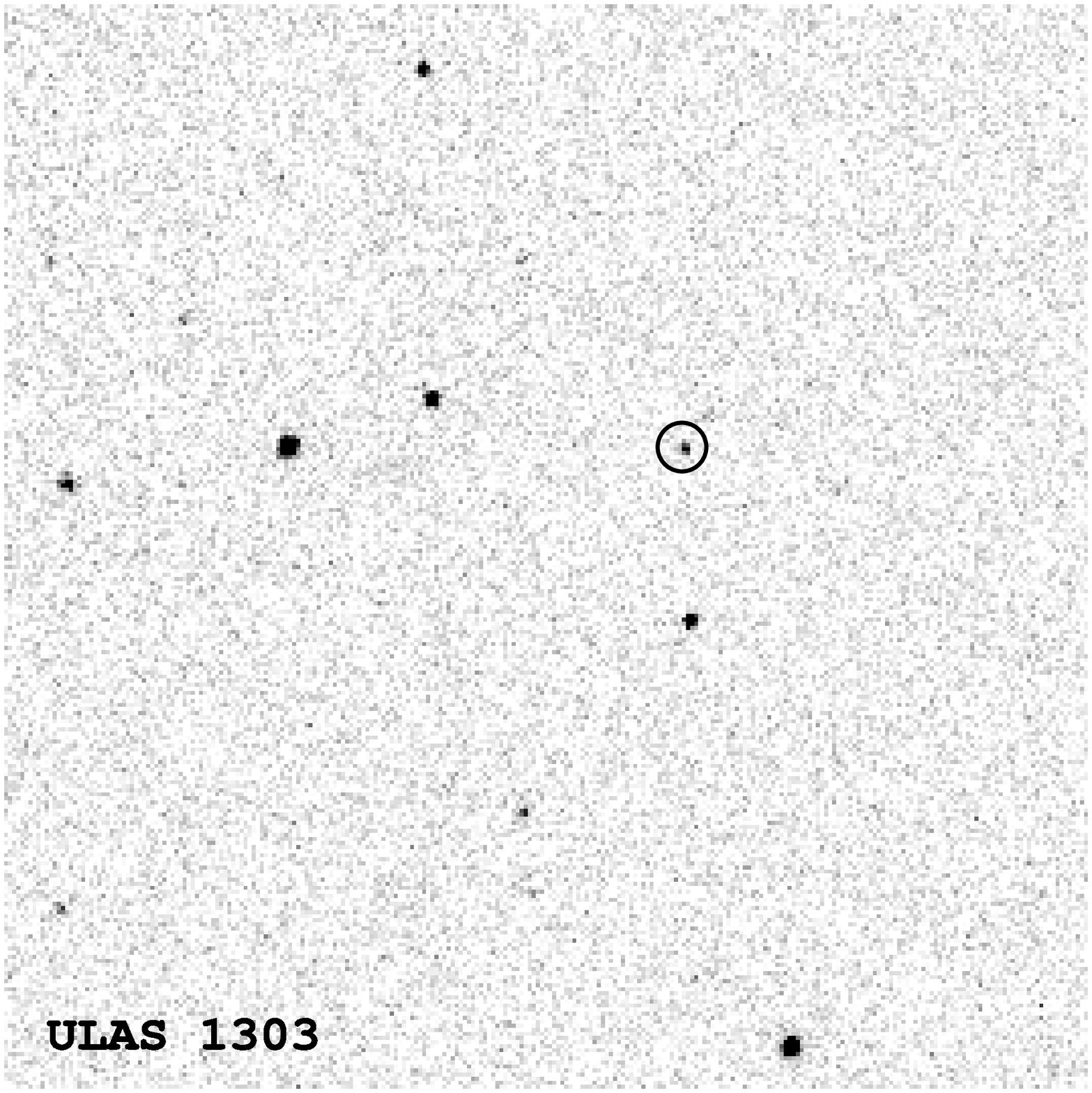} 
\caption{Finding charts based on Omega~2000 methane off images. The fields are 2\,arcmin on each side; North is up and East is left.}
\label{FC}
\end{figure}

For the purpose of our follow-up GROND programme (with scheduled times at the beginning of the night, see subsection\,\ref{GROND}), we requested a right ascension between 7 and 14\,hours.
This returns 19 candidates.
Among those, 13 objects, which have \mbox{$-0.1<J-H<0.0$}, are matched within 3\arcsec\ with a SDSS $griz$ source and have spectral types earlier than M6 \citep{West05}.
Six objects remain and we obtained follow-up data for the four bluest objects, with $J-H<-0.05$, leaving out the two objects with $-0.05<J-H<0$.
The finding charts are given in Fig.\,\ref{FC}.
We did not perform astrometric cuts.

 We perform a rough estimate of the area searched for this work: we count the number of distinct $(6.8\,\arcmin)^2$ footprints (one quarter of a WFCam chip) with at least one stellar-type object with $YJHK$ detections and quality parameters similar to our T~dwarf selection (in terms of photometric errors and quality flags). We subtract the similar number obtained for DR4+ (10881~footprints) from the DR5+ number (14630) and find an area of 186\,sq.deg.
 Because the LAS footprint is mostly contiguous, the (reasonable) choice of a different footprint size and/or origin changes the result by no more than 5\%.

\subsection{GROND follow-up photometry} \label{GROND}

In April 2009 we obtained broad-band $g'r'i'z'JHK$ imaging data of the brown dwarf candidates, using GROND on the ESO/MPG~2.2m telescope located in La~Silla. GROND is a simultaneous imager which uses six dichroics to split the beam into the seven channels \citep{Grein07,Grein08}.
The GROND photometric system is close to SDSS's in the optical and to 2MASS's in the near-infrared. 
The main difference is in the $i$ filter: as the SDSS $riz$ filters overlap at their $\sim 70$\% transmission values, the GROND $i$ is narrower than the SDSS $i$ filter.
However, the central wavelengths are very close.
The colour transformations into the SDSS AB and 2MASS Vega systems of \citet{Grein08} lists negligible colour corrections, except for the $J$~band.
Because we do not have enough observations of {T-type} brown dwarfs to improve the photometric calibration for that spectral class, we report here the photometry in the GROND system, with no colour correction applied.

We observed all targets with the {\tt 20m4td} observation block, with slow read-out mode, four dither positions, 20\,min total integration time in $JHK$ and 24.6\,min in $g'r'i'z'$.
We describe the observations of the four objects reported in this article in Table\,\ref{Tgrond}.

We reduced the data using the GROND pipeline, which is based on the Pyraf/IRAF libraries \citep{Grein08}.
The pipeline corrects the individual images for bias, dark and flat-field. 
It then corrects for geometrical distortions, sky-subtracts the near-IR images, and shifts and co-adds the dithered images.
The source catalogues, positions and aperture photometry are extracted using SExtractor \citep{Berti96}.
Finally, the astrometry is corrected for rotation, pixel scale and translation using the SDSS and 2MASS catalogues as reference, and the photometric zero points can be determined using the same catalogues.

We actually re-determine the zero points, {using a few dozen stars observed by the SDSS and UKIDSS. 
We previously transform the SDSS photometry into the GROND system using the colour transformations of  \citet{Grein08}, and the UKIDSS photometry into the 2MASS system using the transformations of \citet{Hewet06}.}
The typical precision is 5\%, to be added to the SDSS (1--2\%) and UKIDSS (4\%) calibration accuracy. The GROND photometry is listed in Table\,\ref{Ttg}. We do not include the calibration errors in the photometric errors reported in that table.
For non-detections ($gri$ bands), we report the 10-$\sigma$ limit, calculated as the average magnitude of objects with 0.1-mag errors.

We refine the image registration of the GROND co-added images using {\tt scamp} \citep{Berti06}. We use as reference catalogue the UKIDSS positions measured in the band of the closest central wavelength. Bright objects (SNR$>$30, a few dozens) are used to derive a preliminary transformation, which is refined with all objects with SNR$>$3 (one hundred). We use second-order polynomial transformations.
The typical positional dispersion in each direction is 30--70\,mas for the bright star sample, depending on the filters and fields. We add quadratically this dispersion to the photo-noise centring error reported by SExtractor.

\begin{table}
 \centering
 \begin{minipage}{.48\textwidth}
  \caption{Observing logs from GROND and Omega~2000}
  \begin{tabular}{lcccc}
  \hline
    Instrument   & \multicolumn{2}{c}{GROND} & \multicolumn{2}{c}{Omega~2000} \\
    Target           & Date$^1$ &  Seeing$^2$ & Date$^1$ &  Seeing\\
 \hline
    ULAS~J1149$-$0143 & April 29 & 1.5'' & May ~8 & 1.2\arcsec \\
    ULAS~J1153$-$0147 & April 22 & 1.2'' & May 12 & 1.1\arcsec \\
    Ross 458C & April 22 & 1.4'' & May 12 & 1.0\arcsec \\
    ULAS~J1303+1346 & April 30 & 1.9'' & May 12 & 0.9\arcsec \\
 \hline
\multicolumn{5}{l}{$^1$ Year 2009. $^2$~As measured on the $J$-band images.} \\
\label{Tlog}
\label{Tgrond}
\end{tabular}
\end{minipage}
\end{table}

\subsection{Omega2000 methane imaging} \label{O2000cibles}

{The methane imaging is an efficient method to obtain confirmation and preliminary spectral typing of faint T~dwarf candidates using medium-size telescopes.}
Some extragalactic contaminants may mimic T-dwarf methane colours, if their spectrum shows a strong emission line which may fall into the methane off filter transmission for certain redshifts. Quasars have strong H$\alpha$ emission (among other lines) and for redshifts $z=1.30$--1.48 have AB methane colours of $-0.27$ to $-0.73$ (at $z=1.41$). We have indeed detected two known and one new quasars over 1\,sq.deg. (see Section\,\ref{O2000cibles}). However, quasars {at those} redshifts are easily detected in the optical and show optical colours clearly different from those of T~dwarfs.

As the GROND photometry and astrometry supported the brown dwarf hypothesis for these four targets, we obtained DDT observations in the $H$-band methane on and off medium-band filters of the Omega~2000 near-infrared wide-field imager on the Calar Alto 3.5-m telescope \citep{CBJ00}.

In May 2009, we obtained 6-min total-integration time observations in the methane off filter and 25-min integrations in the methane on filters. The observations are described in Table\,\ref{Tlog}. The integrations were split in 1-min co-added images of six 10-sec integrations. Between each co-added image we dithered the telescope to allow for sky and bias subtraction.

The observing set-up and photometric data reduction is similar to those performed to calibrate the Omega~2000 methane filter set. We refer the reader to the appendix\,\ref{methane_cal} for details regarding the method and the data analysis.
The astrometry was derived in a similar way as for the GROND observations, using {\tt SExtractor} and {\tt scamp}, and we refer the reader to Section\,\ref{GROND}.
We find that the dispersion of positions relative to the reference catalogue is minimised for a first-degree polynomial transformation.

\begin{figure}
\includegraphics[width=.48\textwidth]{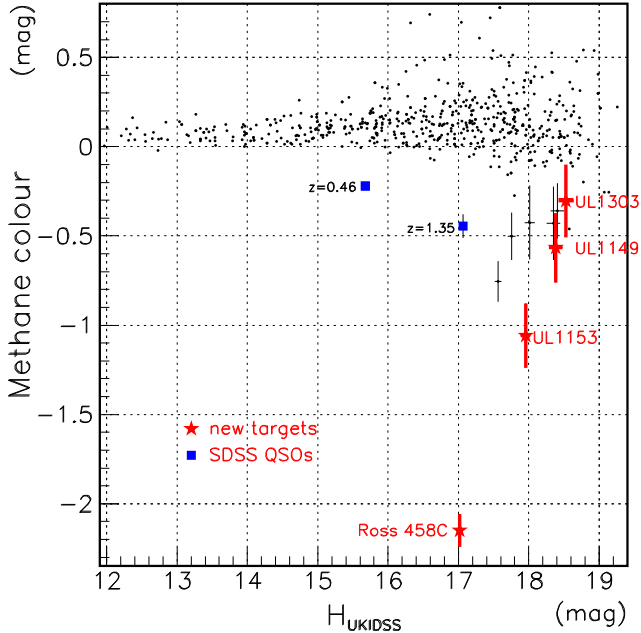}
\caption{Methane colour for our new objects (stars, with statistical error bars), and the field stars in their fields.
We signal two known quasars, LBQS\,1146-0128 with $z=0.46$ and  SDSS\,J130324.21+134045.0 with $z=1.35$. We also indicate the error bars of objects with large, suspicious methane colour. 
Most are artefacts.}
\label{Fmethane2}
\end{figure}

\begin{figure}
\includegraphics[width=.48\textwidth]{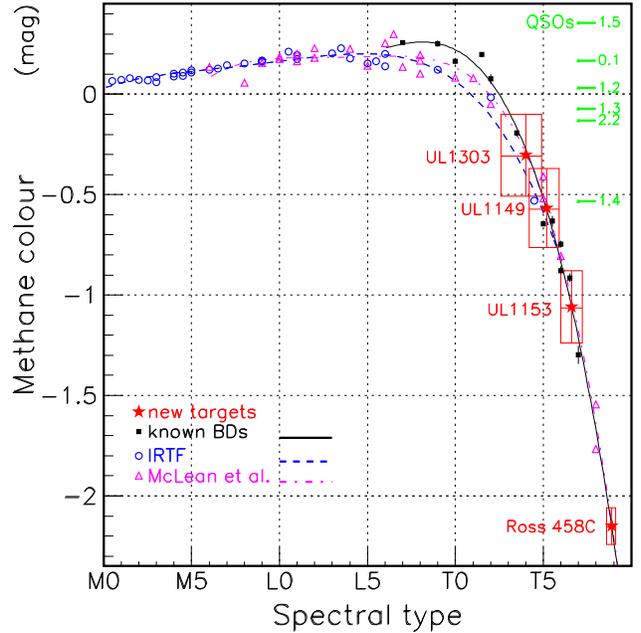}
\caption{Methane colour for our new objects (stars, with statistical error boxes), the reference brown dwarfs (squares, with statistical error bars), and synthetic photometry using the \citet{McLea03} and IRTF spectral libraries (triangles and circles respectively). We superimpose the {polynomial fits for the observed indices (third-degree polynome; in black solid line) and synthetised indices (fourth-degree polynomes; in blue dashed line: IRTF, in magenta dash-dotted line: McLean et al.). See Appendix\,\ref{methane_cal} for details.} We indicate {some} methane colours of quasars with redshifts between 0 and 4.}
\label{Fmethane}
\end{figure}

We report the science targets' methane colour and the spectral type derived from it, along with 1-$\sigma$ error bars, in Table\,\ref{Ttgst}. They are illustrated in Fig.\,\ref{Fmethane2} and \ref{Fmethane}.
The statistical error includes only the photo-noise on the targets' methane colour measurements,
while the systematic errors indicate the calibration errors (see Subsection\,\ref{syserrors}).  

{We derive the photometric distances based on the spectral type and the $J$, $H$ and $K$ magnitudes (see Table\,\ref{Ttgst} and subsection\,\ref{Interpret}). 
}

\subsection{Proper motions}

We use the astrometry derived from the UKIDSS DR5+, GROND and Omega~2000 data to measure the proper motion of our candidates. The UKIDSS~$H$ and $K$ images were obtained in 2007 for Ross~458C or in 2008 for the rest of the sample, the $Y$ and $J$ images in 2008, and the GROND and Omega~2000 images in 2009, resulting in a 1- or 2-year (Ross~458C) baseline.

{We perform the image registration of the Omega~2000 and GROND images using the UKIDSS DR5+ catalogue as the astrometric reference, using the closest filter, in terms of central wavelength.}

We assign an accuracy of 20\,mas for each UKIDSS position. 
{We confirm this rough estimate, originally based on the UKIDSS position residuals, by using the $J$-band second-epoch observations over 42-sq.deg., released in DR6+. 
The relative position root-mean-square  is 33\,mas along the right ascension, and 35\,mas along the declination, for stars with $15<J<19$ and a signal-to-noise ratio larger than 20.
With a time baseline of 2--3\,years, some of the dispersion is due to the real proper motions, and a Besan\c{c}on simulation of those stars estimates this contribution to a small 5\,mas/yr \citep{Robin03}.
Assuming Gaussian, independent noise on each epoch position, the single-frame accuracy in each direction is about 22\,mas for this sample.
}

The GROND and Omega~2000 errors include the photo-noise accuracy and the systematics due to the registration process. The proper motion fit assumes that all errors are independent. 
The systematics introduced by the reference catalogue errors also propagate to all GROND and Omega~2000 measurements.

We report all the measurements for Ross~458C in Table\,\ref{Tpm}, 
and the proper motions of all targets in the same table.  
We give $\mu_\alpha=\frac{d\alpha}{dt}$ rather than the projection of the proper motion along $\alpha$.
{We similarly derived proper motions of the brown dwarfs used to derive the calibration of the Omega2000 methane filter set, and those are presented in section\,\ref{PMcal} and Table\,\ref{Tpmref}.}

\begin{table*}
 \centering
 \begin{minipage}{\textwidth} 
  \caption{Astrometric data on Ross~458AB, of our targets and of the calibration brown dwarfs, as well as the photometric distance and tangential velocity.}
  \begin{tabular}{lllllr@{$\pm$}lr@{$\pm$}lcc}
  \hline
   Name     & Ref. & MJD or &  RA (J2000) & DEC (J2000) & \multicolumn{2}{c}{$\mu_\alpha$} & \multicolumn{2}{c}{$\mu_\delta$} & dist. & $v_\bot$\\ 
                   &          & Epoch  & hms & dms &  \multicolumn{2}{c}{(mas/yr)} & \multicolumn{2}{c}{(mas/yr)} & pc & km/s\\ 
 \hline
   Ross~458AB & (1) & 2000.0 &  13:00:46.582 & +12:22:32.604 & $-618.8$ & 1.8 & $-16.6$ & 1.3 & $11.4\pm 0.2$ & $33.4\pm0.6$ \\ 
   (BD+13$^{\rm o}$2618) & (2) & &  &  &  \multicolumn{2}{c}{$-636.2$} &  \multicolumn{2}{c}{$-25.5$}  \\ 
                                               & (3) & &  &  &  \multicolumn{2}{c}{$-641.2$} &  \multicolumn{2}{c}{$-23.7 $}  \\ 
                        & (1) & 2007.97 &  13 00 46.245 & +12 22 32.47 & \multicolumn{2}{c}{--} & \multicolumn{2}{c}{--} \\ 
 \hline
   Ross~458C & UKIDSS Y & 54572.5 & 13 00 41.729 & +12 21 14.69 \\ 
                          & UKIDSS J & 54572.5 & 13 00 41.730 & +12 21 14.73 \\ 
                          & UKIDSS H & 54224.5 & 13 00 41.771 & +12 21 14.75 \\ 
                          & UKIDSS K & 54224.5 & 13 00 41.771 & +12 21 14.75 \\ 
                          & GROND z & 54943.13 & 13 00 41.698 & +12 21 14.73 \\ 
                          & GROND J & 54943.13 & 13 00 41.688 & +12 21 14.73 \\ 
                          & GROND H & 54943.13 & 13 00 41.696 & +12 21 14.73\\ 
                          & GROND K & 54943.13 & 13 00 41.696 & +12 21 14.65 \\ 
                          & O2000 on & 54960.02 & 13 00 41.683& +12 21 14.72 \\ 
                          & O2000 on & 54963.97 & 13 00 41.683 & +12 21 14.66 \\
                          & O2000 off & 54963.98 & 13 00 41.675 & +12 21 14.81\\ 
                          & fit                & 2007.97 &  13 00 41.743 & +12 21 14.72 & $-629$ & 29 & $-23$ & 26 \\ 
                        & (1) & 2000.0 &  13 00 42.080 & +12 21 14.86 & $-618.8$ & 1.8 & $-16.6$ & 1.3 \\ 
\hline
  ULAS~J1149$-$0143        & --                 & 2008.28 &  11 49 25.584 & $-01$ 57 08.28 & $-51$ & $91$ &  +9 & 85 & 53 & 13 \\ 
  ULAS~J1153$-$0147        & --                 & 2008.29 & 11 53 38.735 & $-01$ 36 25.56 & $-517$ & 73 & $-321$ & 72 & 29 & 85\\
  ULAS~J1303+1346       & --                 & 2008.25 & 13 03 18.281 & +13 56 46.32 & +36 & 76 & $-92$ & 76 & 73 & 34\\
\hline
\multicolumn{5}{l}{(1) \citet{Perry97}. (2) \citet{Zacha03}. (3)  \citet{Ducou06}. }  \\
\label{Tpm}
\end{tabular}
\end{minipage}
\end{table*}

\begin{table*}
 \centering
 \begin{minipage}{\textwidth}
  \caption{Photometry of our targets. The UKIDSS $YJHK$ magnitudes are in the WFCam {\it Vega} system; GROND $iz$ (or 10-$\sigma$ limits) are in the GROND {\it AB} system; the Omega~2000 methane colour is in the Vega system.}
  \begin{tabular}{lccccccccccccl}
   Name     & \multicolumn{2}{c}{ $z_{\rm SDSS}$}& \multicolumn{2}{c}{$Y_{\rm UKIDSS}$}& \multicolumn{2}{c}{$J_{\rm UKIDSS}$} & \multicolumn{2}{c}{$H_{\rm UKIDSS}$} & \multicolumn{2}{c}{$K_{\rm UKIDSS}$} & \multicolumn{2}{c}{spectral type$\pm$(stat)$\pm$(sys)}\\
 \hline
  ULAS~J1149$-$0143   &  \multicolumn{2}{c}{$-$} & \multicolumn{2}{c}{$19.30\pm  0.07$} & \multicolumn{2}{c}{$18.15\pm  0.05$} & \multicolumn{2}{c}{$18.38\pm  0.10$} & \multicolumn{2}{c}{$18.34\pm  0.20$} & \multicolumn{2}{c}{T$5.2^{+0.9}_{-1.2}\pm1$} \\ 
  ULAS~J1153$-$0147   &  \multicolumn{2}{c}{$-$} & \multicolumn{2}{c}{$19.04\pm  0.06$} & \multicolumn{2}{c}{$17.68\pm  0.04$} & \multicolumn{2}{c}{$17.95\pm  0.07$} & \multicolumn{2}{c}{$17.72\pm  0.12$} & \multicolumn{2}{c}{T$6.6\pm 0.8\pm0.3$} \\ 
  Ross~458C                      &  \multicolumn{2}{c}{$20.24\pm 0.19$}              & \multicolumn{2}{c}{$17.72\pm  0.02$} & \multicolumn{2}{c}{$16.69\pm  0.01$} & \multicolumn{2}{c}{$17.01\pm  0.04$} & \multicolumn{2}{c}{$16.90\pm  0.06$} & \multicolumn{2}{c}{T$8.9\pm 0.3\pm0.2$} \\ 
  ULAS~J1303+1346        &   \multicolumn{2}{c}{$-$} & \multicolumn{2}{c}{$19.78\pm  0.08$} & \multicolumn{2}{c}{$18.51\pm  0.04$} & \multicolumn{2}{c}{$18.53\pm  0.11$} & \multicolumn{2}{c}{$18.58\pm  0.18$} & \multicolumn{2}{c}{T$4.0^{+1.3}_{-1.8}\pm1$} \\ 
\hline
   Name     & \multicolumn{2}{c}{$i'_{\rm GROND}$}& \multicolumn{2}{c}{$z'_{\rm GROND}$} & \multicolumn{2}{c}{$J_{\rm GROND}$} & \multicolumn{2}{c}{$H_{\rm GROND}$} & \multicolumn{2}{c}{$K_{\rm GROND}$} &  \multicolumn{2}{c}{}\\
\hline 
   ULAS~J1149$-$0143 & \multicolumn{2}{c}{$>23.12$}& \multicolumn{2}{c}{$22.40\pm0.38$} & \multicolumn{2}{c}{$18.40\pm 0.05$} & \multicolumn{2}{c}{$18.19\pm 0.06$} & \multicolumn{2}{c}{---}\\
   ULAS~J1153$-$0147 & \multicolumn{2}{c}{$>23.16$}& \multicolumn{2}{c}{$21.62\pm 0.27$} & \multicolumn{2}{c}{$17.92\pm 0.05$} & \multicolumn{2}{c}{$17.95\pm 0.05$} & \multicolumn{2}{c}{$18.01\pm 0.10$}\\ 
   Ross~458C                  & \multicolumn{2}{c}{$>21.3$}   & \multicolumn{2}{c}{$20.86\pm 0.20$} & \multicolumn{2}{c}{$16.97\pm  0.03$} & \multicolumn{2}{c}{$17.02\pm  0.03$} & \multicolumn{2}{c}{$16.46\pm  0.05$}\\
   ULAS~J1303+1346   &  \multicolumn{2}{c}{$>22.77$}& \multicolumn{2}{c}{$>22.59$} & \multicolumn{2}{c}{$18.84\pm 0.06$} & \multicolumn{2}{c}{$19.20\pm 0.10$} & \multicolumn{2}{c}{$19.33\pm 0.20$}\\ 
\hline
   Name     &&  \multicolumn{2}{c}{$z_{\rm GROND}-Y$} & \multicolumn{2}{c}{$(Y-J)_{\rm UKIDSS}$} & \multicolumn{2}{c}{$(J-H)_{\rm UKIDSS}$} & \multicolumn{2}{c}{$(H-K)_{\rm UKIDSS}$} &  & \multicolumn{2}{c}{CH$_4$ colour} \\ 
\hline
    ULAS~J1149$-$0143  & & \multicolumn{2}{c}{$1.10\pm 0.38$} & \multicolumn{2}{c}{$+1.15\pm  0.08$} & \multicolumn{2}{c}{$-0.23\pm  0.11$} & \multicolumn{2}{c}{$+0.04\pm  0.22$} & & \multicolumn{2}{c}{$-0.57\pm 0.20$}  \\ 
    ULAS~J1153$-$0147  & & \multicolumn{2}{c}{$2.63\pm 0.27$} & \multicolumn{2}{c}{$+1.36\pm  0.07$} & \multicolumn{2}{c}{$-0.27\pm  0.08$} & \multicolumn{2}{c}{$+0.23\pm  0.14$} & & \multicolumn{2}{c}{$-1.06\pm 0.18$}  \\  
    Ross~458C                    & & \multicolumn{2}{c}{$3.08\pm 0.19$} & \multicolumn{2}{c}{$+1.03\pm  0.03$} & \multicolumn{2}{c}{$-0.33\pm  0.04$} & \multicolumn{2}{c}{$+0.09\pm  0.07$} & & \multicolumn{2}{c}{$-2.15\pm 0.11$}  \\  
    ULAS~J1303+1346      & & \multicolumn{2}{c}{$>2.81$} & \multicolumn{2}{c}{$+1.27\pm  0.09$} & \multicolumn{2}{c}{$-0.02\pm  0.12$} & \multicolumn{2}{c}{$-0.05\pm  0.21$} & & \multicolumn{2}{c}{$-0.30\pm 0.20$}  \\  
\hline
\label{Ttg}
\label{Ttgst}
\end{tabular}
\end{minipage}
\end{table*}

\section[]{Companionship to Ross~458} \label{companion}

Several independent measurements of the proper motion of Ross~458AB (also DT~Vir, BD+13$^{\rm o}$2618) are available in the literature. We quote a selection in Table\,\ref{Tpm}. Hereafter we use the Hipparcos measurement for simplicity; in most cases the conclusions are identical for all values.

The proper motions of Ross~458AB and Ross~458C agree within 0.5$\sigma$ in both directions.
Ross~458AB is detected in the Luyten Half-Second catalogue (LHS), with a proper motion of 0.714\,\arcsec/yr \citep{Luyte79}, and is bright enough to be included in the Hipparcos catalogue \citep{Perry97}.
There are 2120 stars with proper motions larger than 0.6\,\arcsec/yr in the LHS catalogue, or one every 20\,sq.deg. Among those there are 639~objects with an Hipparcos parallax measurement, and 308 have a distance between 5 and 20\,pc, compatible with the photometric parallax of Ross~458C. The number of such objects over a 5-arcmin-radius disk is $2.10^{-4}$.

Ignoring the distance constraints, but considering the small proper motion difference: there are 443 stars with proper motion within 1$\sigma$ of Ross~458C's proper motion in the LHS catalogue \citep{Luyte79}. (At the brightness of Ross~458AB, the LHS catalogue is close to complete.) In addition, the proper motion direction of Ross~458C is known to $\pm 2\deg$ (1$\sigma$), so that we expect 5~stars over the sky to share its proper motion within the uncertainties, or $3.10^{-6}$ over the 5-\arcmin-radius disk area.
Searching for stars with proper motions within $5\sigma$ (to increase the statistics) of Ross~458C's in \citet{Salim03}, which covers three quarters of the sky, returns 143~stars, or 7.6~stars for a $1\sigma$ interval over the whole sky.

\begin{figure}
\includegraphics[width=.48\textwidth]{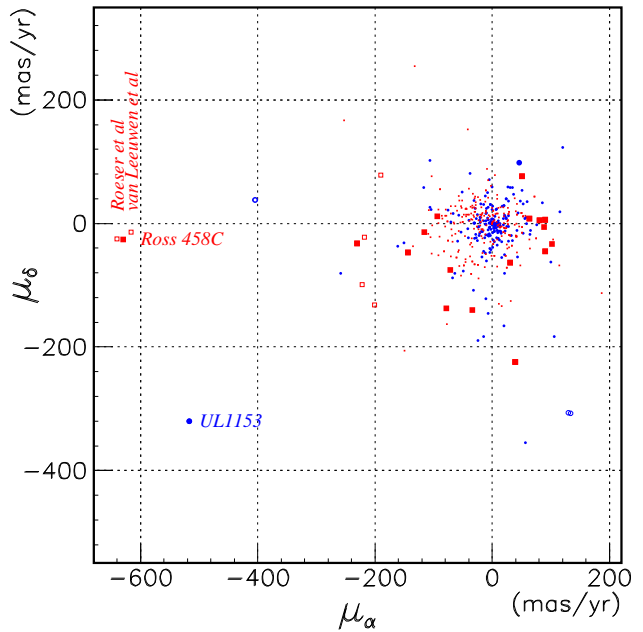}
\caption{Measured proper motion vectors of the 15-arcmin field objects around Ross~458 (squares) and ULAS~J1153 (circles); large symbols show significant proper motions at the 3-$\sigma$ level. Open symbols show catalogued stars with proper motions larger than 0.2\arcsec/yr, within $1\deg$ of the targets (from \citet{vLeeu07} and \citet{Roese08}, see text).}
\label{Fpm}
\end{figure}

Fig.\,\ref{Fpm} illustrates the rare match of the proper motions of Ross~458AB and our target. We show the proper motions we measure for all field objects around Ross~458 and ULAS~J1153, as well as the high proper motions (\mbox{$\mu>0.2\,$\arcsec/yr}) from \citet{vLeeu07} and \citet{Roese08} (other proper motion catalogues give similar results) within $1\deg$ of the targets.

Either way, the probability for Ross~458AB and Ross~458C to be chance neighbours is negligible.

\section[]{Properties of the new brown dwarfs} \label{prop}

\subsection[]{Ross~458C}

\subsubsection[]{Properties of Ross~458AB}

Ross~458AB is a binary made of a M7 dwarf orbiting a M0.5 dwarf (although early references report M2, or even M3), with a 14.5-yr period  \citep{Heint94}. The separation was 0.475\arcsec\ in February 2000, or 5.4\,AU  \citep{Beuzi04}. However, the latter authors point out an inconsistency between their measurements and the results of  \citet{Heint94}.

Ross~458A is very active, as indicated by its H$\alpha$ emission strength and variability (see Table\,\ref{TpropAB}). \citet{West08} estimate an age of $400\pm400$\,Myr for M1~active dwarfs.
Its rotational velocity of 10\,km/s means it is not tidally-locked, also implying an age smaller than $\sim 1$\,Gyr.

The space velocity presented by \citet{Monte01} makes it a possible member of the Hyades supercluster, which has a space velocity of $(U,V,W)=(-40,-17,-3)$\,km/s \citep{Zucke04}.
Considering other published proper motions does not change this conclusion.
The age of the Hyades is estimated at $(648\pm45)$\,Myr \citep{DeGen09}. However the members of its moving group show a dispersion in age (0.4--2\,Gyr) and metallicity.
Possibly the common motion should be attributed to tidal effects of the massive rotating bar, rather than a common birthplace.
The metallicity of the cluster of $\rm [Fe/H]=+0.14$ is compatible with the solar metallicity estimate of \citet{Alons96}.
We note, however, that \citet{Hawle97} report a radial velocity of $-43.4$\,km/s, leading to $W=-40$\,km/s, which would be incompatible with a Hyades moving group membership.

Taking into account these various constraints we estimate the age of Ross~458AB to less than 1\,Gyr.
We summarise the properties of Ross~458AB in Table\,\ref{TpropAB}.

\begin{table*}
 \centering
 \begin{minipage}{\textwidth}
  \caption{Description of Ross~458AB.}
  \begin{tabular}{lcr@{$\pm$}lll}
  \hline
    Parameter   & Component & \multicolumn{2}{c}{Value} & Unit& Reference \\
  \hline
    Type              & A & \multicolumn{2}{c}{M0.5} & & \citet{Hawle97} \\ 
                          & B & \multicolumn{2}{c}{M7} & & \citet{Beuzi04} \\
     Mass & A & \multicolumn{2}{c}{0.6} & $\rm M_\odot$ & \citet{Legge92} \\
                 & B & \multicolumn{2}{c}{0.06--0.09} & $\rm M_\odot$ & \citet{Baraf98} \\
     Age & A & \multicolumn{2}{c}{$\leq 1$} & Gyr \\ 
     distance & AB & 11.4 & 0.2 & pc & \citet{vLeeu07} \\
     $M_{\rm bol}$ & AB & \multicolumn{2}{c}{7.90} & mag & \citet{Moral08} \\
  \hline
     X (0.5--2.0\,keV) & AB & \multicolumn{2}{c}{$975.10^{-17}$} & W/m2   & \citet{Fisch00} \\
     $B$              & AB & \multicolumn{2}{c}{11.206} & mag & \citet{Perry97}\\
     $V$              & AB & \multicolumn{2}{c}{9.52} & mag & \citet{Hawle97} \\
     $V$              & AB & 9.758 & 0.026 & mag & \citet{Kharc07}\\ 
                          & B & \multicolumn{2}{c}{$>13.2$} & mag & \citet{Henry99} \\ 
                          & AB & \multicolumn{2}{c}{9.73} & mag & \citet{Besse90} \\ 
     $R$              & AB & 8.81 & 0.026 & mag &??? \\ 
     $R$              & AB &  \multicolumn{2}{c}{8.75} & mag & \citet{Besse90} \\ 
     $I$              & AB &  \multicolumn{2}{c}{7.67} & mag & \citet{Besse90} \\ 
     $J$               & AB & 6.437 & 0.021 & mag & \citet{Cutri03} \\
     $H$            & AB &  5.786 & 	0.017& mag & \citet{Cutri03} \\
     $H$ (1.62$\,\mu$m) & AB & \multicolumn{2}{c}{5.75} & mag & \citet{Morel78} \\
     $K$	           & AB & 5.578 & 0.016 & mag & \citet{Cutri03} \\
         		           & B &  \multicolumn{2}{c}{9.99} & mag & \citet{Beuzi04} \\
     $L$  (3.5$\,\mu$m) & AB & \multicolumn{2}{c}{5.40} & mag & \citet{Morel78} \\
 \hline
     Variability & AB & \multicolumn{2}{c}{$V=0.05$} & mag & \citet{Pojma05} \\
                        &        & \multicolumn{2}{c}{$Hp=0.11$} & mag & \citet{Pojma05} \\
     TiO5 & AB & \multicolumn{2}{c}{0.700} & \AA & \citet{Hawle97,Moral08} \\
     H$\alpha$ EW & AB & \multicolumn{2}{c}{1.660} & \AA & \citet{Hawle97,Moral08} \\
                          &         & \multicolumn{2}{c}{1.4--2.1} & \AA & \citet{Short98} \\
     $\log g$ & AB & \multicolumn{2}{c}{5.00} & & \citet{Alons96} \\
     $\rm [Fe/H]$ & AB & \multicolumn{2}{c}{0.00} && \citet{Alons96}  \\
     $T_{\rm eff}$& AB & \multicolumn{2}{c}{3870} & K & \citet{Alons96}  \\
     $v.\sin i$& AB & \multicolumn{2}{c}{10} & km/s & \citet{Short98} \\
 \hline
     Rad.vel. & AB & $-11.23$ & 0.11 & km/s &  \citet{Nidev02} \\ 
     $U$	& AB & $-29.74$ & 1.00& km/s &  \citet{Monte01} \\
     $V$ & AB & $-18.54$ & 1.03& km/s &  \citet{Monte01} \\
     $W$ & AB & $-8.85$ & 4.83& km/s &  \citet{Monte01} \\
\hline
     Period & & \multicolumn{2}{c}{14.5} & years & \citet{Heint94} \\
     Separation & & \multicolumn{2}{c}{5} & A.U. & \citet{Beuzi04} \\
\hline
\label{TpropAB}
\end{tabular}
\end{minipage}
\end{table*}

\subsubsection[]{Interpretation} \label{Interpret}

\begin{figure}
\includegraphics[width=.48\textwidth]{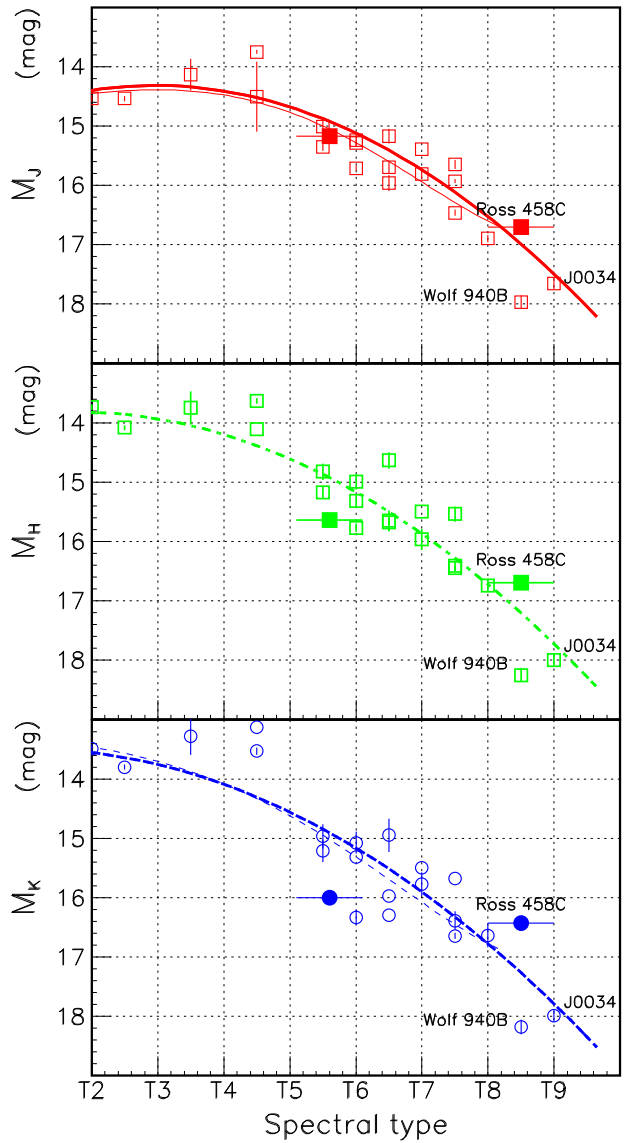}
\caption{
Magnitude (in the 2MASS system) vs. spectral type diagram for dwarfs later than T2, excluding known close binaries. We correct the UKIDSS magnitudes into the 2MASS system (see Appendix\,\ref{uk2m}). 
We overplot a tentative $2^{\rm nd}$ polynomial fit to the data, 
as well as the previous determination by \citet{Loope08bi} in $J$ and $K$, as thin lines.
}
\label{Fcmd}
\end{figure}

Given the parallax of Ross~458AB, we deduce the absolute magnitudes of Ross~458C to be: $M_J=16.40\pm0.04$ and $M_K=16.61\pm0.06$ in the WFCam system.
{We convert the UKIDSS photometry into the 2MASS system using the colour transformations we derived in section\,\ref {uk2m}., and place in Fig.\ref{Fcmd}} 
our target in the absolute magnitude vs. spectral type diagram, {along with} 21~known brown dwarfs with trigonometric parallax (excluding known close binaries). 
The spread of magnitudes for a given spectral type is large in the sample, certainly due to variations in the gravity (i.e. mass/age), unresolved binaries, and to a lesser extend metallicity.
We nevertheless fit the {2MASS photometry} with a second-degree polynomial:
\begin{equation}
   M_J\:=15.08-0.515\times n+0.0871\times n^2\;\rm mag
\end{equation}
\begin{equation}
   M_H=14.03-0.253\times n+0.0738\times n^2\;\rm mag
\end{equation}
\begin{equation}
   M_K=13.56-0.138\times n+0.0676\times n^2\;\rm mag
\end{equation}
where $n$ is now the T spectral sub-type (e.g. $n=0$ for T0). 
{It is clear that this function is not based on even the simplest modeling of the brown dwarf structure.
The reduced $\chi^2$s are 56, 35 and 32 in the $J$, $H$ and $K$ bands respectively (the photometry is less accurate for the longer wavelengths), and the dispersions are 0.41\,mag, 0.50\,mag and 0.66\,mag respectively. (In addition, the data are heterogeneous, with errors on the absolute magnitude ranging for 3--4\% to 15--20\%.)
A proper relation between absolute magnitude and spectral type would take into account additional parameters, especially age.
}
We use those relationships below to derive (rough) photometric distances of our isolated targets.

Ross~348C's absolute photometry would favour a T8 ($J$ band) or even T7.5 ($K$ band) spectral type.

With a similar spectral type (T8.5) and metallicity ($\rm [Fe/H]=-0.06\pm0.20$), the much redder $J-K$ colour of Ross~458C compared to that of Wolf~940B is a clear indication of its lower gravity, which produces a reduced H$_2$ collision-induced absorption in the $K$ band.

The $z-J$, $J-H$ and $H-K$ colours of Ross~458C are also within $1\sigma$ of those of ULAS~J003402.77$-$005206.7, while Ross~348C is 0.3\,mag redder in $Y-J$ than ULAS~J0034 (but it is only a $3\sigma$ effect). 
{The recent parallax of ULAS~J0034 by \citet{Smart10} gives a distance modulus of $0.21\pm0.11$\,mag larger than that of Ross~458C, so that the latter is $1.25\pm0.12$\,mag brighter in absolute magnitude than ULAS~J0034.} Binarity (with a nearly equal-mass companion), a slightly earlier spectral type (e.g. T8.5, still compatible with identical colours), and/or a {significantly} larger distance to ULAS~J0034  (e.g. 17\,pc, at $1\sigma$) could explain the discrepancy.

Alternatively, it could be due to the difference in age and mass (and radius). \citet{Legge09} has used Spitzer IRS 7.6--14.5$\mu$m spectroscopy, IRAC photometry as well as near-infrared spectroscopy and detailed modeling of ULAS~J0034 to better constraints its parameters. They find a degeneracy between (higher) metallicity and (lower) gravity on the 2.2 and 4.5$\mu$m fluxes, but for the Solar metallicity of Ross~458AB and a distance of 16\,pc, they favour $T_{\rm eff}=600$\,K, $g=100\rm\,m.s^{-2}$, $R=0.12\rm\,R_\odot$, $M=$5--8\,M$_{\rm Jup}$ and an age of 0.1--0.2\,Gyr.
This is again a strong indication of the youth of Ross~458ABC, and the very small mass of Ross~458C.

\begin{figure}
\includegraphics[width=.48\textwidth]{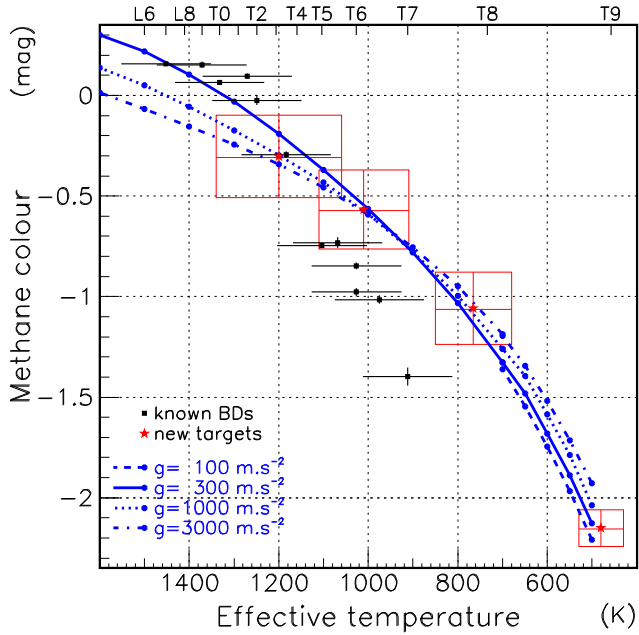}
\caption{Methane colour for our new objects (stars, with error boxes), and synthetic photometry using the ``no clouds'' atmospheric models of \citet{Saumo08}, for various gravities. 
We calculate the effective temperature of the reference brown dwarfs using the effective temperature vs. spectral type relationship of \citet{Steph09}, which has a residual dispersion of $\approx 100$\,K (squares, with error bars).
}
\label{FmethaneMM}
\end{figure}

The spectral type we derive from the methane colour is extrapolated for spectral types later than T7, so the T$8.9\pm 0.3$ classification we obtain for Ross~458C may have to be corrected. We use atmospheric models by \citet{Saumo08} to calculate synthetic methane colour as a function of effective temperatures, for various gravities, to lower effective temperatures than those of our reference brown dwarfs. We plot the results in Fig.\ref{FmethaneMM}.

We report the effective temperatures and methane colours for our known brown dwarfs, using the relation~(6) of \citet{Steph09}. 
We note that the effective temperatures of our mid- and late-T dwarfs seem offset compared to the models by $\sim 100\,$K. The offset is even larger for the mid-T dwarfs when using the relationship of \citet{Golim04}.
{The methane line lists are known to be incomplete, which may explain the discrepancy for the mid-T dwarfs. For instance, the methane colour derived from the best model fit of SDSS\,J111009.99+011613.0, a red T5.5 dwarf, by \citet{Steph09} is bluer than what the observed spectrum indicates.}

\citet{Liu07} reports effective temperatures for two T7.5 of $\sim 800$\,K, i.e. 100\,K cooler than what \citet{Golim04} or  \citet{Steph09} predict for the T7 SDSS\,J162838.77+230821.1. 
On the other hand, \citet{dBurg09}, comparing high resolution spectroscopy with models, derive for four T7--T8~dwarfs temperatures systematically larger than 800--900\,K (1\,$\sigma$).

We therefore do not attempt to derive an effective temperature for our targets based on their methane colours. Instead, {we use the model colours in Fig.\,\ref{FmethaneMM}} to first conclude on the rather low dependency of  the methane colour to the gravity: from 0\,mag at 1000\.K to 0.3\,mag at 500\,K, over 1.3\,orders of gravity. 
This dependency is larger than our photometric errors for Ross~458C, but would not affect the effective temperature determination by more than 50\,K for any object cooler than 1100\,K. 

Secondly we can use the synthetic photometry to consistently estimate the range of mass and gravity based on the evolutionary models of \citet{Saumo08}. For a model effective temperature of 500\,K, and an age between 0.1 and 1\,Gyr, their Fig.4 predicts for Ross~458C a gravity of 50 to 300\,m.s$^{-2}$, and a mass between 5 and 14\,Jupiter masses.

\subsubsection[]{Binding energy}

From the spectral type of Ross~458A, we derive a mass of $0.6\rm\,M_\odot$ \citep{Legge92}. The contribution of Ross~458B and C to the total mass of the system is negligible. At the current projected separation of $r\sin i=1168$\,A.U., corresponding to 102\arcsec, the binding energy is:
\begin{equation}
   E=4.0.10^{42} \times \sin i \times \frac{m_{\rm Ross 458C}}{0.02\rm M_\odot}\rm erg
\end{equation}
where $i$ is the orbit inclination with respect to the line of sight.

\begin{figure} 
\includegraphics[width=.48\textwidth]{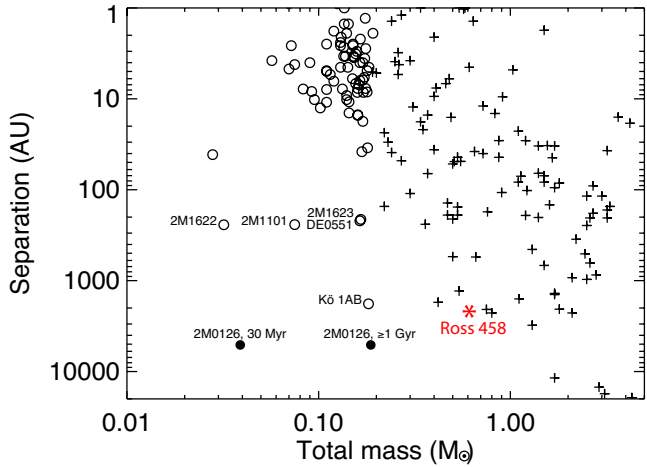}
\caption{Separation of binaries as a function of total mass, from \citet{Artig07}. The star represents Ross~458ABC.
}
\label{Fsepar}
\end{figure}

In Fig.\ref{Fsepar} we reproduce the distribution of binary separation vs. total mass as compiled by \citet{Artig07}, with Ross~458ABC. This system has a large separation compared to the main locus of binaries.
{\citet{Faher10} compiled a more complete list of wide binaries and discuss the distribution of binding energies. There are a dozen systems with a similar total mass and smaller binding energies.
These authors also discuss the more frequent high-order multiplicity in weakly bound systems, which would help them survive disruption compared to simple binaries. In fact, Ross~458 is a triple system, and the overluminosity of Ross~458C could be an indication of binarity. 
}

\subsection[]{Other targets}

\subsubsection[]{ULAS~J1149$-$0143}

ULAS~J1149$-$0143 has a significant methane colour of $\-0.57\pm0.20$. It is not detected in SDSS and must be a cool object. Its methane colour is more that $3\sigma$ away from those of M and L~dwarfs. We estimate a spectral type of T$5\pm1.5$.

With a one-year baseline, we cannot measure the proper motion of ULAS~J1149$-$0143, but place an upper-limit of $\sim 0.3$\arcsec/yr, compatible with a distance as small as 20\,pc and a tangential velocity as large as 30\,km/s. The photometric distance of a T5 dwarf with ULAS~J1149$-$0143's $J$ and $K$ magnitudes would be $\sim 50$\,pc.

\subsubsection[]{ULAS~J1153$-$0147} \label{J1153}

ULAS~J1153$-$0147 has a significant methane colour of $\-1.06\pm0.18$, $6\sigma$ away from those of M and L~dwarfs. It is not detected in SDSS. We estimate a spectral type of T$6.5\pm1$.

We measure the (large) proper motion of ULAS~J1153$-$0147 at $7\sigma$. It is also 
the latest and the brightest of our targets after Ross~458C. It has a photometric distance of $\sim 29$\,pc, and a tangential velocity of $\sim 85$\,km/s. 
{We ran a Besan\c{c}on simulation of the field to estimate what fraction of stars with kinematics similar to ULAS~J1153$-$0147.
It has a velocity along $V$ of $-50$\,km/s to $-82$\,km/s for a radial velocity between $-20$\,km/s and $+30$\,km/s (66\% of all stars have their radial velocity in this range). 
The model predicts that 60\% of stars with such $V$ velocities belong to the thick disk (this fraction decreases to 46\% for a radial velocity range corresponding to 90\% of stars in the field).
The age dispersion of T-type dwarfs in the Solar neighbourhood is expected to differ from that of field stars \citep{Allen05}, and therefore their velocity distributions may differ as well, but this gives an indication that ULAS~J1153$-$0147 may be an old object.
}

There are no objects with a similar proper motion in our data, and none within 30\,arcmin in Simbad.

\subsubsection[]{ULAS~J1303+1346}

ULAS~J1303+1346 has the weakest methane colour of our four targets. Taken at face value, it is only 2$\sigma$ away from the stellar population, and therefore has a $\sim 2.5\%$ chance not to be a T~dwarf.
However, among the $\sim 387$ field stars, only six have a larger methane colour (1.6\%), all detected in SDSS $riz$ bands. (Among those six two are classified as galaxies by SDSS, two have $ugr$ colours typical of quasars, and the last two are classified as stars.)
Therefore, the probability for a given object to have a spurious (i.e. without methane absorption) methane colour of $-0.4$\,mag in this data set is $<2.5\%$.
In addition, we selected the target because of its very red $Y-J$ and blue $J-H$ colours, hardly compatible with a stellar or a low-redshift quasar colour, and its UKIDSS stellar profile, incompatible for most galaxies of that brightness. Hence most contaminants are excluded.
This does not prove its sub-stellar nature {\it beyond any doubt}, but it is very likely.
With a one-year baseline, we fail to detect the proper motion of ULAS~J1303+1346, but as for ULAS~J1149$-$0143 this does not exclude that it is a brown dwarf.

\section{Conclusions}

We have searched $\approx$186\,sq.deg. of UKIDSS Large Area Survey newly released in DR5+.
Using simple colour and quality selection criteria, we selected four candidates for fast photometric follow-up with GROND on the ESO/MPG~2.2m and Omega~2000 on the Calar Alto 3.5-m telescope.
We present a calibration of the Omega~2000 methane filter set.
Methane imaging confirms three T~dwarfs and strongly supports the brown dwarf hypothesis of the fourth target.
The latest-type target, Ross~458C, a T8-9 brown dwarf, is a companion to a M0.5+M6 binary which provides us with an accurate distance measurement  {(11.4\,pc)} and a rough age estimate {(less than 1\,Gyr)}.
The absolute magnitudes and $J-K$ colour point to a very young age, or possible binarity.
{Evolutionary models indicate a mass close to the deuterium burning minimum mass.}
This object will become an important well-characterised example of the latest T~dwarfs.
{In addition we identify by proper motion and photometric distances two thick disk candidates, ULAS~J1153$-$0147 and SDSS~J1540+3742 {(section\,\ref{PMcal})}. }

\section*{Acknowledgments}

We thank Calar Alto Observatory for allocation of director's discretionary time to this programme and the observatory staff for its expeditious observations. We also thank Sascha Quanz for his precious comments, Steve Boudreault, Mark Marley, Annie Robin, and Markus Schmalz for their advice, and the referee Ben Burningham for his suggestions and corrections and his quick revision.

Part of the funding for GROND (both hardware as well as personnel) was generously granted from the Leibniz-Prize to Prof. G. Hasinger (DFG grant HA 1850/28--1).

This research has made use of the {\sc Simbad} database, operated at C.D.S., Strasbourg, France,
       and of the M, L, and T dwarf compendium housed at {\tt DwarfArchives.org} and maintained by C.\,Gelino, D.\,Kirkpatrick, and A.\,Burgasser.
       
       The UKIDSS project is defined in \citet{Lawre07}. UKIDSS uses the UKIRT Wide Field Camera \citep[WFCAM; ][]{Casal07}. The photometric system is described in \citet{Hewet06}, and the calibration is described in \citet{Hodgk09}. The pipeline processing and science archive are described in \citet{Hambl08}. We thank the WSA support team for their help.       
    
       Funding for the SDSS and SDSS-II has been provided by the Alfred P. Sloan Foundation, the Participating Institutions, the National Science Foundation, the U.S. Department of Energy, the National Aeronautics and Space Administration, the Japanese Monbukagakusho, the Max Planck Society, and the Higher Education Funding Council for England. The SDSS Web Site is {\tt http://www.sdss.org/}.

{This publication makes use of data products from the Two Micron All Sky Survey, which is a joint project of the University of Massachusetts and the Infrared Processing and Analysis Center/California Institute of Technology, funded by the National Aeronautics and Space Administration and the National Science Foundation.}

\appendix

\section[]{Omega2000 methane filter set calibration} \label{methane_cal}

\subsection{Observations}

To calibrate our methane filter set, in June 2008 we observed 12 reference brown dwarfs with known spectral type, derived from spectroscopy using the \citet{Burga06cl} classification scheme. The spectral types vary from L7- to T7. We selected the targets (see Table\,\ref{Tmethane} for their full names) based on their brightness, observability, and to obtain a good coverage of the late L and T spectral types.
Here and through-out this article, we use the classification scheme of \citet{Burga06cl} whenever available.

\begin{figure}
\includegraphics[width=.48\textwidth]{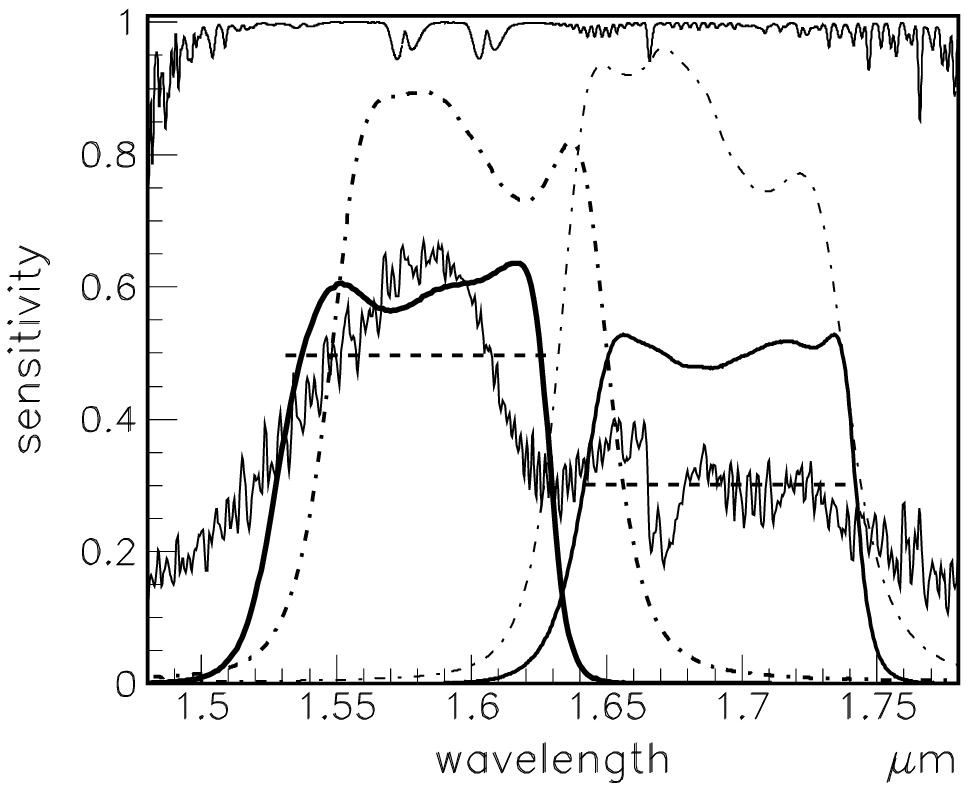}
\caption{Complete filter sensitivity profile, including blocking filter and detector sensitivity (thick solid lines). The spectrum of the T5 2MASS~J05591914-1404488 is overplotted \citep{Cushi05}, its synthetic photometry (dashed lines), as well as the profiles of the AAT filters (only) used by \citet{Tinne05}, for comparison (dot-dashed lines), and a typical atmospheric transmission (solid line, top). Note that some measurements were provided by the manufacturer and corrected to the operating conditions.}
\label{Ffilters}
\end{figure}

Observations were performed with the Omega~2000 instrument, mounted on the primary focus of the 3.5-m telescope at the Calar Alto Observatory in Spain. The filter profiles, including a blocking filter and the Omega2000 Hawaii~2 detection sensitivity, are shown in Fig.\,\ref{Ffilters}. The filters are well separated and well positioned with respect to the methane absorption feature, although some water absorption affects the methane off filter. Dithered exposures of around one minute were taken in the two filters of interest, methane off and on, and also in the $H$ band. 
Exposures varied in number, due to observational constraints, but are always more numerous for the methane on filter, taking into account the lower expected flux in this band. Atmospheric conditions over the four nights of observation were good, with seeing varying from 0.9 to 1.2\arcsec. Calibration frames such as dark exposures and dome or dusk flatfields were taken each night. 

Preliminary treatment such as the construction of a bad-pixel map, flatfields for the two filters and a dark current image was performed using standard IRAF routines. Individual frame treatment, flatfielding, dark current- and bad pixel corrections as well as realignment and summation were carried out using the Omega 2000 data reduction pipeline, written by Ren\'e Fassbender. 

A first astrometry has been computed using {\tt koords}\footnote{Part of the Karma toolkit: \newline \tt www.atnf.csiro.au/computing/software/karma/} and the 2MASS catalog as reference. Photometric measurements were performed using SExtractor, over the large sample of the field stars and object. Since only relative photometry was of interest, no zero-point determination has been necessary. Depending on the extraction threshold, we observed wide dispersion for the faintest objects; hence we determined the zero-methane colour using a sample arbitrarily restricted to the 50 brightest objects. No distinction has been made between stars and galaxies for the determination of the zero-colour, as explained in section 2.3.

One of the targets, the T2.5 dwarf 2MASS~J1546+4932, appeared closely blended to another star, because of its proper motion shift, and was rejected.

The results are presented in Table\,\ref{Tmethane} and Fig.\,\ref{Fmethane}.  We discuss the calibration of our photometry below in subsection\,\ref{calcal}.

\begin{table*}

 \begin{minipage}{\textwidth} 
  \caption{Calibration sample for the Omega~2000 methane filter set}
  \begin{tabular}{@{}llllrrcr@{$\pm$}l}
  \hline
      Name & \multicolumn{1}{c}{$H_{\rm 2MASS}$} & Spectral & Date$^1$  & \multicolumn{2}{c}{Exp. time (min)} & Seeing & \multicolumn{2}{c}{Methane} \\
   &  & type &   & Off & On & (arcsec) &  \multicolumn{2}{c}{colour} \\
 \hline
SDSSp  J134646.45-003150.4  &	15.46$\pm$0.11 		 &T6.5	   & Jun 20    & 6  & 15 & 1.1 &	$-0.915$ & 0.033	\\ 
2MASS  J15031961+2525196    &	13.86$\pm$0.03  &		T5	   & Jun 20    & 6  & 15 & 1.1 &	$-0.645$ & $0.013$	\\ 
SDSS   J152039.82+354619.8  &	14.58$\pm$0.05  		&T0$\pm$1   & Jun 20    & 6  & 9  & 1.1 &	+0.165& 0.011		\\
SDSS   J152103.24+013142.7  &	15.58$\pm$0.10  		&T2	   & Jun 20    & 6  & 12 & 1.2 &	$+0.976$ & 0.023	\\
2MASSI J1526140+204341        &	14.50$\pm$0.04  &	 L7$^2$	   & Jun 20    & 6  & 15 & 1.1 &	+0.256 & 0.010		\\
SDSS   J154009.36+374230.3  &	15.35$\pm$0.10 		 &L9$\pm$1.5 & Jun 20    & 6  & 9  & 1.1 &	+0.253 & 0.016		\\
2MASS  J15461461+4932114    &	15.13$\pm$0.09  &	T2.5$\pm$1 & Jun 20    & 6  & 12 & 1.0 &	 \multicolumn{2}{c}{}		\\
2MASS  J21392676+0220226    &	14.17$\pm$0.05  &	T1.5	   & Jun 18-19 & 6  & 12 & 1.0 &	+0.197 & 0.012		\\
2MASS  J21543318+5942187    &	15.77$\pm$0.17  &	T6	   & Jun 19-20 & 10 & 25 & 1.1 &	$-0.746$ & 0.017	\\
SDSSp  J162414.37+002915.6  &	15.52$\pm$0.10	 &T6	   & Jun 21    & 10 & 25 & 0.9 &	$-0.878$& 0.021	\\
SDSS   J162838.77+230821.1  &	16.11$\pm$0.15 		 &T7	   & Jun 21    & 10 & 25 & 0.9 &	$-1.297$ & 0.044	\\
SDSS   J163022.92+081822.0  &	16.33$\pm$0.29 		 &T5.5	   & Jun 18    & 10 & 25 & 1.1 &	 $-0.632$ & 0.027	\\
SDSSp  J175032.96+175903.9  &	15.95$\pm$0.13  	&T3.5	   & Jun 20    & 10 & 20 & 1.0 &	$-0.194$ & 0.015	\\
\hline
\multicolumn{9}{l}{$^1$ Year 2008. $^2$ Optical classification \citep{Kirkp00}.} \\
\label{Tmethane}
\end{tabular}
\end{minipage}
\end{table*}

\subsection{Synthetic photometry}

To complement the observations described above we use the measured transmission profile of the methane filter set at 77\,K, the operating temperature, to predict the synthetic colours of F to T7 dwarfs. We use the spectral libraries of IRTF/SpeX\footnote{http://irtfweb.ifa.hawaii.edu/$\sim$spex/IRTF\_Spectral\_Library/} \citep{Cushi05} and of Keck/NIRSPEC\footnote{http://www.astro.ucla.edu/$\sim$mclean/BDSSarchive/} \citep{McLea03}, which both provide calibrated $F_\nu$ spectra, with brown dwarfs as late as the T4.5 dwarf 2MASS~J05591914$-$1404488 and the T8 2MASS~J04151954$-$0935066 respectively.

We calculate the expected counts in each band and derive the methane colour in the AB system
{(we calculate the offset into the Vega system below)}.
We find the methane colour to be insensitive to luminosity class (I--V) at the few mmag level.

\subsection{Photometric calibration} \label{calcal}

We present our colour measurements $H_{\rm off}-H_{\rm on}$ {and the colours vs. spectral type relationships} in the Vega system, with A0-type stars having a null colour.

\subsubsection{Synthetic photometry}

\citet{Tinne05} found a null colour for stars earlier than K. As the earliest dwarfs in \citet{Cushi05} are F~stars, we use the AB photometry of those to correct the synthetic AB photometry into the Vega system. We find an average AB colour of $(-0.202\pm 0.003)$\,mag with a dispersion of 1.1\%.

The results are superimposed in Fig.\,\ref{Fmethane}. We fit a fourth-order polynomial function on both datasets independently:
\begin{equation}
   C_m=(-16411 + 1106 n - 4.67 n^2 - 9.33 n^3 - 1.211 n^4).10^{-5}
\end{equation}
\begin{equation}
   C_m=(1783+43.0 n - 21.26 n^2 + 5.361 n^3 - 0.4126 n^4). 10^{-4}
\end{equation}
where $n$ is the spectral type: $n=-10$ for M0, $n=0$ for L0, $n=10$ for T0, etc, 
for the IRTF/SpeX (valid for $n\leq16$) and Keck/NIRSPEC, respectively.
The dispersions are 29\,mmag and 66\,mmag respectively.

\subsubsection{Observations}

In order to save telescope time under photometric conditions, we do not wish to observe spectro-photometric standard stars and perform a full calibration. Instead, we use the methane colours of the field stars, observed under the same seeing and transparency conditions, to correct the raw photometry into the Vega system.
Stars have few absorption features in the $H$ band, and \citet{Tinne05} found that the methane colour in the AAT/IRIS2 system is a smooth, monotonic function of spectral type, and near-infrared colours, between the A and the early~L spectral classes, with a peak-to-peak variation of less than 0.2\,mag. 

\citet{Tinne05} use their relationship between methane colour and near-infrared colours for A to M stars to derive the zero point. As we do not have observed methane colours for stars, we would have to rely on the synthetic photometry to perform the corrections. Instead, we estimate the mean colour of field objects and assign it a fix value.

Our sensitivity to stars with small methane colours is much larger in the $H_{\rm on}$ filter{, because of the longer integration time in that filter}. To avoid the Eddington bias which would lead to the overestimate of $H_{\rm off}$ fluxes of objects detected a low signal-to-noise ratios in the $H_{\rm off}$ band and large signal-to-noise ratios in the $H_{\rm on}$ band, we only consider bright enough sources. 

 The problem of the distinction between stars and galaxies arises, but we do not expect the latter to show any methane absorption feature. 
 Emission-line galaxies may however show non-zero methane colours; galaxies of various types will also show a limited range of colours as B--M stars do.
 Possible aperture photometry biases due to the measurement of extended sources should not change from one filter to the next, but a different seeing between the two integrations may introduce systematics. As we limit ourselves to the brightest objects in the field, this eliminates most galaxies.

We run a Besan\c{c}on model stellar simulation \citep{Robin03} of the field of Ross~458, whose stellar population, with a Galactic latitude of $75\deg$, is typical of those of the UKIDSS--LAS fields. Combining the population simulation  with the synthetic methane photometry of A--M stars, we obtain a mean methane AB colour of $-0.10$\,mag (corresponding to a K star), largely insensitive to the magnitude upper-limit.

{The procedure is then to subtract from the target raw colour the mean raw colour of stars in the field, add $-0.10$\,mag to obtain the target AB colour, and subtract $-0.20$\,mag to derive the target Vega colour.}

We fit a third-order polynomial function to the spectral type $n$ (\mbox{$n=-10$} for M0, $n=0$ for L0, $n=10$ for T0):
\begin{equation}
   C_m= 0.273-0.0769\times n + 0.0183\times n^2-1.11.10^{-3}\times n^3
\end{equation}
We ignore the photometric errors in the fit as some of them seem to be underestimated.
{The {residual} dispersion is 82\,mmag.
}

\begin{table*}
 \centering
 \begin{minipage}{\textwidth} 
  \caption{Proper motion and astrometry of the calibration brown dwarfs, as well as the photometric distance and tangential velocity.}
  \begin{tabular}{llllr@{$\pm$}lr@{$\pm$}lcc}
  \hline
   Name     & MJD or &  RA (J2000) & DEC (J2000) & \multicolumn{2}{c}{$\mu_\alpha$} & \multicolumn{2}{c}{$\mu_\delta$} & dist. & $v_\bot$\\ 
                    & Epoch  & hms & dms &  \multicolumn{2}{c}{(mas/yr)} & \multicolumn{2}{c}{(mas/yr)} & pc & km/s\\ 
\hline
SDSSp  J1346-0031  & 2004.706 & 13 46 46.252 & -00 31 50.88 & $-505$ & 14 & $-105$ & 14 & 14.6$^1$ & 35 \\ 
2MASS  J1503+2525    &  2004.508 & 15 03 19.641 & +25 25 22.51 & +75 & 12 & +560 & 17 & 7 & 19 \\ 
SDSS   J1520+3546  & 2003.312 & 15 20 39.828 & +35 46 19.84 & +308 & 12 & -374 & 13 & 13 & 27 \\ 
SDSS   J1521+0131  &  2007.552 & 15 21 03.165 & +01 31 43.11 & $-219$ & 28 & +59 & 32 & 24 & 26 \\ 
2MASSI J1526+2043      & 2000.205 & 15 26 14.005 & +20 43 40.48 & $-227$ & 12 & $-358$ & 13 & 6 & 12 \\ 
SDSS   J1540+3742  & 2004.339 & 15 40 09.344 & +37 42 29.99 & $-226$ & 14 & $-382$ & 15 & 16$^2$ & 44\\ 
2MASS  J1546+4932    &  2006.690 & 15 46 14.829 & +49 32 13.51 & $-40$ & 34 & $+731$ & 17 & 19 & 67\\ 
2MASS  J2139+0220    &  2003.256 & 21 39 26.858 & +02 20 23.09 & +490 & 15 & +144 & 16 & 13 & 31 \\ 
2MASS  J2154+5942   &  2007.254 & 21 54 33.186 & +59 42 19.00 & $-1$ & 18 & $+31$ & 18 & 13 & 2\\ 
SDSSp  J1624+0029  &  2002.773 & 16 24 14.275 & +00 29 15.77 & -388 & 11 & +0 & 11 & 11$^3$ & 20 \\ 
SDSS   J1628+2308  & 2006.084 & 16 28 38.854 & +23 08 19.98 & $+418$ & 21 & $-438$ & 21 & 12 & 34 \\
SDSS   J1630+0818  & 2007.345 & 16 30 22.909 & +08 18 21.57 & $-76$ & 38 & $-87$ & 37 & 18 & 10 \\
SDSSp  J1750+1759  & 2007.822 & 17 50 33.048 & +17 59 04.79 & +183 & 33 & +62 & 33 & 27$^3$ & 24  \\ 
\hline
\multicolumn{10}{l}{$^1$ \citet{Tinne03}. $^2$ \citet{Loope08bi} give a photometric distance of 22\,pc, within $2\,\sigma$ of our value.} \\
\multicolumn{10}{l}{We use their estimate as our relationship does not apply for L dwarfs. $^3$ \citet{Vrba04}.}  \\
\label{Tpmref}
\end{tabular}
\end{minipage}
\end{table*}

Fig.\,\ref{Fmethane} shows the good agreement between the predictions and the observations for T~dwarfs later than T4. This gives us confidence that we can derive a precise spectral type based on our methane colour. The agreement is not so good for L/T transition brown dwarfs.

{\subsection{Systematic uncertainties} \label{syserrors}

The errors on the spectral type derived from the methane colour are the sum of the photometric errors on the methane colour measurement for each target, and of systematic errors resulting from the photometric errors and spectral type errors of the reference brown dwarfs and the possibly inappropriate fitted function. We can use the dispersion around the fitted functions to estimate the systematic errors. 
The dispersion for the observed methane colours is larger than that derived from the synthetic photometry, so we conservatively use it hereafter.
Because the slope of the relationship becomes steeper for later objects, the systematic error decreases with spectral type, from 0.29\,subtype at T5 to 0.16\,subtype at T8.

The dispersion, however, does not reproduce all the contributions to the systematic error, because our reference sample is small, and may not be representative, in terms of e.g. gravity or dust composition, of our target sample (we note however that both samples are magnitude-limited). 
From Fig.\,\ref{FmethaneMM}, we see however that the sensitivity of the method to the gravity is modest for spectral types later than T6, at about 0.2\,subtype over a range of $g=100$--$3000\,\rm m.s^{-2}$, to 1\,subtype at T4, and much more for earlier types.
Because our reference sample includes objects of various gravities, this contribution is already partly taken into account in the fit dispersion, but we conservatively add it to the systematic errors.
}

\subsection{Proper motions and companions} \label{PMcal}

We use the discovery positions and our Omega~2000 positions to derive the positions of all objects in the Omega~2000 field of view. We perform the image registration of the Omega~2000 data using the discovery catalogue as the astrometric reference, using the same technique as for the targets (see Section\,\ref{GROND}).
The results are presented in Table\,\ref{Tpmref}.
We obtain a good agreement with published values, such as those of \citet{Tinne03}. 
However our values significantly differ from those of \citet{James08} for 2MASS~J1503+2525 and SDSS~J1521+0131.

All the transverse velocities we measure are typical of thin disk velocities, except those of 2MASS~J1546+4932 
and SDSS~J1540+3742.
The former has a positive velocity along the Galactic rotation direction $V$ for a wide range of radial velocities, and therefore belong to the thin disk.
As in section \ref{J1153}, a Besan\c{c}on simulation of the field to estimate what fraction of stars with kinematics similar to the latter.
SDSS~J1540+3742 has a velocity along $V$ of $-80$\,km/s to $-43$\,km/s for a radial velocity between $-50$\,km/s and $+20$\,km/s (63\% of all stars have their radial velocity in this range). 
The model predicts that 63\% of stars with such $V$ velocities belong to the thick disk (this fraction decreases slightly for a larger range of radial velocities).
It is therefore possibly a thick disk object, but it is not possible to assign it with certainty to the thick disk population based on our data.

We also searched for other T~dwarfs in the field of view, using the methane colour. We found none, but we identified an object, SDSS~J163023.89+081230.9, with a colour corresponding to a T5 dwarf, but with $ugriz$ SDSS detections. Because of this and of its optical colours, it is most likely a $z\approx 1.4$ quasar, in a side area covered by SDSS but not by the SDSS QSO survey (Fan, priv. com.).

\section[]{UKIDSS--2MASS colour transformations} \label{uk2m}

\begin{figure} 
\includegraphics[width=.48\textwidth]{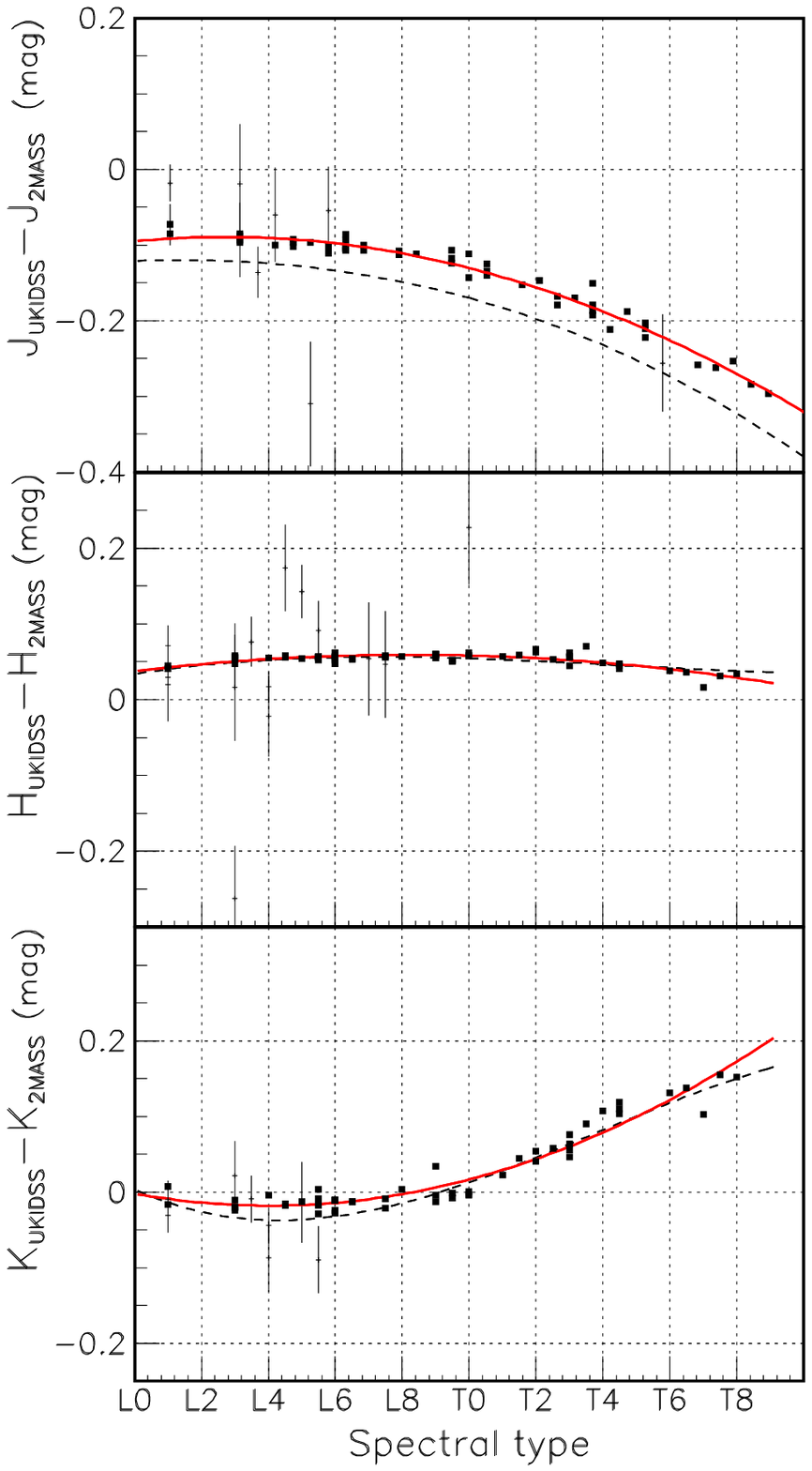}
\caption{UKIDSS--2MASS colours as a function of spectral type, for 52 ultracool dwarfs \citep[square, from][]{Hewet06}, { as red solid lines}. 
We superimpose the observed colours of ultra-cool dwarfs with 2MASS detections, observed with UKIDSS (with error bars).
We superimpose the transformations between the MKO and 2MASS systems of  \citet{Steph04} as black dashed lines.
}
\label{Fuk2m}
\end{figure}

We have used the UKIDSS and 2MASS magnitudes calculated by \citet{Hewet06} of 52 ultracool dwarfs (22 T~dwarfs) to fit second-order polynomials of the \mbox{UKIDSS--2MASS} colours as a function of (infrared) spectral types (see Fig.\,\ref{Fuk2m}). We obtain:
\begin{equation}
   M_J=-0.136-0.0127\times n-0.00085\times n^2\;\rm mag
\end{equation}
\begin{equation}
   M_H=+0.058-0.0011\times n-0.00031\times n^2\;\rm mag
\end{equation}
\begin{equation}
   M_K=+0.017+0.0116\times n+0.00097\times n^2\;\rm mag
\end{equation}
where $n$ is the T spectral sub-type (e.g. $n=0$ for T0). 
The dispersions are 1.0\%, 0.6\% and 1.5\% in the $J$, $H$ and $K$ band respectively.

We searched the UKIDSS DR6 Large Area Survey and Galactic Cluster Survey catalogues for UKIDSS magnitudes of ultracool dwarfs detected by 2MASS, within 5\arcsec.
Unfortunately, 2MASS measurements of T-type dwarfs with good accuracy ($<0.2$\,mag) are scarce, especially at longer wavelengths.
Removing the worst outlier and subtracting the 2MASS and UKIDSS photometric errors, we find a dispersion of 6\%, 9\% and 1\% in the $J$, $H$ and $K$ bands respectively, based on 9, 14 and 8 objects.

{Finally we compare our transformations with those that \citet{Steph04} obtained between the 2MASS and MKO systems \citep[the WFCam filter set was designed to match the MKO system, ][]{Hewet06}.
The agreement is excellent except in the $J$ band, for which we measure $J_{\rm UKIDSS}-J_{\rm 2MASS}$ colours 3\% (5\%) larger for L- (T-, resp.) type dwarfs than the $J_{\rm MKO}-J_{\rm 2MASS}$ colours of \citet{Steph04}.
}

\bibliographystyle{mn2e}
\bibliography{mybib.bib}


\label{lastpage}

\end{document}